\documentclass[aps, prb, reprint, amsmath, amssymb, groupedaddress, longbibliography]{revtex4-2}
\usepackage[colorlinks=true, allcolors=blue]{hyperref}
\usepackage{graphicx}
\usepackage{xcolor}

\newcommand{\lt}{\left}
\newcommand{\rt}{\right}
\newcommand{\pa}{\partial}
\newcommand{\bS}{\mathbf{S}}
\newcommand{\bk}{\mathbf{k}}
\newcommand{\bq}{\mathbf{q}}
\newcommand{\bx}{\mathbf{x}}
\newcommand{\bG}{\mathbf{G}}
\newcommand{\bg}{\mathbf{g}}
\newcommand{\ba}{\mathbf{a}}

\begin{document}

\title{Non-linear spin wave theory in the strong easy-axis limit of the triangular XXZ model}

\author{Achille Mauri}
\author{Siebe Roose}
\altaffiliation{Present address: Helmholtz-Zentrum Berlin f\"{u}r Materialien und Energie,
Hahn-Meitner-Platz 1, 14109 Berlin, Germany}

\author{Fr\'{e}d\'{e}ric Mila}

\affiliation{
Institute of Physics, Ecole Polytechnique F\'{e}d\'{e}rale de Lausanne (EPFL), CH-1015 Lausanne,
Switzerland}

\date{\today}

\begin{abstract}
Motivated by recent experimental studies, we investigate the spectrum of the nearest-neighbour
triangular XXZ model within the $1/S$ expansion, in the limit in which the exchange couplings
present a strong easy-axis anisotropy $J_{xy}/J_{zz} \ll 1$.
We show that in the limit in which $1/S \to 0$ and $J_{xy} \to 0$ at fixed $V = J_{zz}/(S J_{xy})$,
the triangular spin model can be reduced to an effective boson model with quartic interactions on
the honeycomb lattice.
This effective model interpolates between a spin-wave ($V \to 0$) and a strong-coupling
limit ($V \to \infty$) and encodes in a simple framework the regimes discussed by Kleine~\emph{et
al.}~[Z. Phys. B Condens. Matter~{\bf 86}, 405 (1992);~{\bf 87}, 103 (1992)].
For zero field, the classical ground state of the model presents an accidental degeneracy, which can
be traced to a simple symmetry of the classical energy.
The model thus offers a transparent realization of a theory with quantum order-by-disorder and a
pseudo-Goldstone mode.
We analyze the spectrum at zero magnetic field by calculating the self-energy at one-loop order.
In the calculation, we introduce a self-consistent renormalization of the energy scale and of the
pseudo-Goldstone energy gap; the latter renormalization is essential to remove infrared divergences
in the on-shell corrections to the energy dispersion.
Finally, we discuss qualitatively the structure of the one-loop corrections in comparison with the
spectrum observed experimentally in K$_{2}$Co(SeO$_{3}$)$_{2}$.
\end{abstract}

\maketitle

\section{Introduction}

The nearest-neighbour XXZ model on the triangular lattice is one of the key models in the theory
of frustrated magnetism, and has attracted extensive interest for decades.
In an early theoretical investigation, Fazekas and Anderson~\cite{fazekas_pm_1974} suggested that
the model may exhibit a quantum spin-liquid ground state in the limit of strong easy-axis
anisotropy $J_{xy} \ll J_{zz}$.
The original arguments in favor of a magnetically disordered ground state, however, were shown to be
inconclusive, and it was later proposed that the system exhibits an ordered ground state, even in
the limit $J_{xy}/J_{zz} \to 0$ which was initially considered favorable for
a disordered state~\cite{kleine_zpb_1992, kleine_zpb_1992b}.
In subsequent analyses, the ground state of the easy-axis XXZ model has been studied extensively by
complementary methods, both in the context of magnetic systems and in the related context of lattice
boson models~\cite{murthy_prb_1997, wessel_prl_2005, melko_prl_2005, heidarian_prl_2005,
burkov_prb_2005, sen_prl_2008, wang_prl_2009, heidarian_prl_2010, zhang_prb_2011, yamamoto_prl_2014,
ulaga_prb_2025, zhu_npj_2025, gallegos_prl_2025, ulaga_arxiv_2025}.
The semiclassical (large-$S$) mean field theory~\cite{miyashita_jpsj_1985, kleine_zpb_1992,
kleine_zpb_1992b, murthy_prb_1997} indicates that the model presents at zero field a canted coplanar
magnetic order, with three-sublattice unit cell.
In the language of lattice boson models, this canted coplanar phase corresponds to a supersolid,
which breaks simultaneously lattice translation and $U(1)$ symmetries~\cite{wessel_prl_2005,
melko_prl_2005, heidarian_prl_2005}.
In the presence of a longitudinal field, parallel to the easy axis the mean-field solution predicts
a sequence of transitions between the low field ``spin-supersolid'' state, a ``solid'' phase
(a $1/3$ magnetization plateau), and a second canted state, before full saturation.

The quantum $S = 1/2$ XXZ model has been analyzed beyond the semiclassical approximation by
complementary methods, including variational theories~\cite{wang_prl_2009, heidarian_prl_2010,
bose_prb_2025}, stochastic series expansion~\cite{zhang_prb_2011}, cluster mean-field
theory~\cite{yamamoto_prl_2014}, tensor-network methods~\cite{gao_npj_2022, xu_prb_2025},  exact
diagonalization~\cite{ulaga_prb_2024, ulaga_prb_2025, ulaga_arxiv_2025}, and density-matrix
renormalization group (DMRG)~\cite{gallegos_prl_2025, ulaga_arxiv_2025}.
In the low-field regime, and for the fully antiferromagnetic model with $J_{xy} > 0$, $J_{zz}>0$,
the results of cluster mean-field theory, stochastic series expansion, and DMRG indicate a
spin-supersolid phase, consistent with the semiclassical prediction.
However, recent exact diagonalization studies~\cite{ulaga_prb_2024, ulaga_prb_2025,
ulaga_arxiv_2025} indicated that the ground state for $S = 1/2$ may be a normal ``solid'' with no
superfluid component.
In addition, the tensor network results reported in Ref.~\cite{xu_prb_2025} indicated a very small
superfluid density, which could be compatible with the absence of superfluidity in the exact ground
state.

The limit $J_{xy}/J_{zz} \to 0$ plays a special role in the model.
In this limit, the theory behaves as a constrained ``quantum Ising model''~\cite{moessner_prb_2001}.
In addition, in this limit, the cases with antiferromagnetic interaction and with ferromagnetic
$J_{xy} < 0$ are equivalent and can be mapped into each other by a unitary
transformation~\cite{wang_prl_2009}.
This implies that the quantum Ising model emerging for small $J_{xy}$, and the related quantum dimer
model~\cite{fazekas_pm_1974, wang_prl_2009} have no sign problem.
Recently, this fact was used in order to perform a quantum Monte Carlo simulation of the dimer
model~\cite{zhu_npj_2025}.
The unitary equivalence between the antiferromagnetic and the ferromagnetic models, combined
with the Matsubara-Matsuda mapping between spins and hard-core bosons, implies also a direct
connection between the antiferromagnetic XXZ spin model and the problem of bosons with unfrustrated
hopping on the triangular lattice.
This model was studied by quantum Monte Carlo in Refs.~\cite{wessel_prl_2005, melko_prl_2005,
heidarian_prl_2005}.
These studies found supersolid ground states.
However, the superfluid stiffness is found to be small in the large $J_{zz}$
limit~\cite{burkov_prb_2005}.

Recently, the limit $J_{xy} \to 0$ of the XXZ model has stimulated a renewed interest, after new
experimental studies of magnetic compounds which realize effectively spin-$1/2$ models on the
triangular lattice~\cite{gallegos_prl_2025, kamiya_nc_2018, gao_npj_2022, arh_nm_2022,
chen_nc_2026, zhu_prl_2024, zhu_npj_2025, jia_prr_2024, chi_prb_2024, ulaga_prb_2025,
ulaga_prb_2024, xu_prb_2025, sheng_innov_2025, woodland_arxiv_2025, ulaga_arxiv_2025}.
In the present work, we are motivated in particular by recent studies of the cobaltite
K$_{2}$Co(SeO$_{3}$)$_{2}$ (in the following ``KCSO'')~\cite{zhong_prm_2020, zhu_prl_2024,
chen_nc_2026, zhu_npj_2025, xu_prb_2025, ulaga_prb_2025, ulaga_prb_2024, zhang_arxiv_2025,
bose_prb_2025, jia_prr_2024, ulaga_arxiv_2025}.

The analyses in Refs.~\cite{zhong_prm_2020, zhu_prl_2024, chen_nc_2026, zhu_npj_2025} show that
this compounds realizes the $S = 1/2$ XXZ model, and has an extremely strong easy-axis anisotropy,
with $J_{xy}/J_{zz} \simeq 0.07$.
Among other observations, inelastic neutron scattering experiments revealed that the dispersion of
the excitations of the compound present deep roton-like minima at the $M$ points of the Brillouin
zone.
These mimima are absent in the linear-spin wave approximation~\cite{zhu_npj_2025, zhu_prl_2024}, and
require an explanation beyond the simple picture of noninteracting magnons.
In addition, Refs.~\cite{zhu_npj_2025, chen_nc_2026, zhu_prl_2024} reported a significant
spectral broadening of the excitations.
Furthermore, Ref.~\cite{zhu_npj_2025} identified the presence of a low-energy excitation,
which was interpreted as a pseudo-Goldstone mode.

In Ref.~\cite{zhu_npj_2025}, the spectrum of KCSO was analyzed by quantum Monte Carlo
simulations of the quantum dimer model associated with the XXZ model in the $J_{xy}/J_{zz} \to 0$
limit.
The results successfully reproduced the emergence of a dispersion minimum at low field, as well as
other salient features of the neutron spectrum at finite field.
A roton minimum was also found in Ref.~\cite{xu_prb_2025}, by tensor-network methods
based on projected entangled-pair states.
In addition, Refs.~\cite{ulaga_prb_2025, ulaga_arxiv_2025} investigated by exact diagonalization
(ED) methods both the triangular XXZ model and an effective nearest-neighbour XXZ model on a
honeycomb sublattice, which provides an approximation to the full triangular spin model.
Ref.~\cite{ulaga_prb_2025} showed that the spectrum of the effective honeycomb model presents a
change of
behavior and features not captured by LSWT at zero magnetization $M^{z} = 0$ and small ratio
$J_{xy}/J_{zz}$.
The analysis of Ref.~\cite{ulaga_arxiv_2025} indicated qualitative differences between the
triangular-lattice XXZ model and the approximation of a honeycomb XXZ model with only
nearest-neighbour interactions.
In particular, the results of Ref.~\cite{ulaga_arxiv_2025} provided indications of a vanishing
superfluid density for the triangular model, and of a finite superfluid density for the honeycomb
model.

In a broader context, anomalous spin excitation spectra and deviations from LSWT have been
investigated in connection with a variety of triangular compounds, such as
Ba$_{3}$CoSb$_{2}$O$_{9}$~\cite{ma_prl_2016, macdougal_prb_2020, chi_prl_2022},
Na$_{2}$BaCo(PO$_{4}$)$_{2}$~\cite{gao_prb_2024, jia_prr_2024, gao_npj_2022, chi_prb_2024}, and
KYbSe$_{2}$~\cite{scheie_np_2024}, which have various degrees of exchange anisotropy, from easy
plane to easy axis.
In addition, series expansion studies have reported a roton-like minimum in the triangular
Heisenberg model~\cite{zheng_prl_2006, zheng_prb_2006}.

While numerical simulations have been reported the nature of the interpretation of the excitation
spectrum of KCSO and, more generally, of triangular compounds continues to pose open questions.
In the case of the triangular Heisenberg model, it has been suggested that, despite the presence of
long-range order, excitations of moderate energy should be more naturally described in terms of
spinon excitations~\cite{zheng_prb_2006, zheng_prl_2006}.
However, a conflicting interpretation was developed, involving only magnon-magnon interactions.
In particular, it was shown that $1/S$ corrections to linear spin-wave theory produce a flattening
of the dispersion relation.
Thus, it was suggested that the roton-like minima observed by series expansion methods may be
accounted in terms of spin-wave interactions (see Refs.~\cite{chernyshev_prb_2009, zheng_prb_2006}
for discussions).
A further mechanism for roton-like minima, spin-wave magnon interactions, was proposed in the case
of square-lattice quantum magnets.
Refs.~\cite{powalski_prl_2015, powalski_spp_2018} developed a theory attributing the roton to a
level repulsion between single-magnon states and three-magnon states, the latter forming resonances
due to interactions.

Following experimental observations on triangular compounds, recent works have developed parton
theories to describe spectral properties of anisotropic XXZ models.
In particular, Ref.~\cite{jia_prr_2024} developed a fermionic parton theory as an explanation for
neutron spectrum observed in Na$_{2}$BaCo(PO$_{4}$)$_{2}$.
This explanation has been discussed as a possible scenario to explain also the observations in
KCSO~\cite{zhu_npj_2025, zhu_prl_2024, jia_prr_2024}.
A recent work~\cite{bose_prb_2025} analyzed a fermionic spinon theory, including a resummation of
the two-spinon correlation function to describe magnon excitations.
Although the analysis of Ref.~\cite{bose_prb_2025} assumed a nearly isotropic exchange
interaction, it was suggested to be applicable qualitatively also to anisotropic XXZ compounds.

A natural question however remains: is it possible that the spectrum of the easy-axis XXZ model can
be explained naturally in terms of magnon-magnon interactions, as it was proposed for the Heisenberg
case?
Can an explanation of this type account for the spectrum of KCSO, in which roton-like minima have
been found to be very pronounced?

The objective of this work is to investigate this scenario, by analyzing spin-wave corrections
in the strongly-anisotropic XXZ model.
When compared to the Heisenberg model, the strongly-anisotropic XXZ model presents important
difficulties.
The first is that in the limit $J_{xy}/J_{zz} \to 0$, the Heisenberg interaction acts effectively
as a constraint.
Refs.~\cite{kleine_zpb_1992} indicated that, because of this, the model becomes strongly coupled,
even for large $S$.
A second difference is that the XXZ model presents an accidental symmetry in the semiclassical
approximation.
This gives rise to a pseudo-Goldstone mode, which is gapless in the linear-spin-wave approximation,
and which acquires a gap due to fluctuations~\cite{murthy_prb_1997, rau_prl_2018}.
The pseudo-Goldstone mode, instead, is a ``true'' Goldstone mode in the Heisenberg case.
The impact of pseudo-Goldstone fluctuations thus introduces new questions, in comparison with the
Heisenberg case.

In this work we take a step towards understanding the renormalization of the single-magnon spectrum
in the triangular XXZ model with $J_{xy} \ll J_{zz}$.
We first show that the usual nonlinear spin-wave expansion can be considerably simplified by
taking the limit $S \to \infty$, $\alpha \to 0$ at fixed $V = 1/(S \alpha)$.
We show that, in this limit the problem becomes equivalent to a soft-core boson model with a
nearest-neighbour repulsion of magnitude $V = 1/(S \alpha)$.
The spin-wave expansion corresponds to the limit $V \to 0$, in which the model behaves
semiclassically, and the interaction $V$ plays the role of the coupling constant, controlling the
loop expansion.
For $J_{xy}/J_{zz} \ll 1/S$ the model is eventually governed by the strong coupling limit $V \to
\infty$.
The fact that $V$ plays the role of the coupling, thus, recovers the prediction that spin-wave
theory is controlled by $J_{zz}/(J_{xy} S)$, as predicted in Ref.~\cite{kleine_zpb_1992}.

In this work, we investigate the first-order corrections in the region of small $V$.
The effective honeycomb-lattice model which we consider exhibits a classical degeneracy, with an
associated pseudo-Goldstone mode, which is a remnant of the pseudo-Goldstone mode of the full
triangular XXZ model.
By analyzing the first-order corrections, we provide an analysis, in a particular model of how
spin-wave renormalizations affect spectral corrections at arbitrary momentum $\bk$ in a model with
pseudo-Goldstone modes.
Our analysis, from this point of view, extends earlier theories which calculated the
pseudo-Goldstone gap at zero momentum~\cite{murthy_prb_1997, rau_prl_2018}.
As we show, the pseudo-Goldstone fluctuations introduce infrared divergences at all momenta in two
dimensions, and must be regularized by introducing a gap renormalization.

Due to the classical degeneracy and the pseudo-Goldstone mode, the corrections which we find
present a complex and singular dependence on $V$ for $V$ small.
The gap of the pseudo-Goldstone mode grows asymptotically as $3.05 \sqrt{V} S J_{xy}$ for $V \to
0$, whereas corrections at finite $\bk$ have a parametrically weaker dependence.
The rapid growth of the pseudo-Goldstone gap with $V$ implies that when $V \simeq 0.5$ the spectrum
has already been reconstructed quite strongly.
Thus we find that values of $V$ of the order of $0.5$ are to be considered moderately large.

In the analysis which we present the strong anisotropy $J_{xy}/J_{zz} \to 0$ provides a first
advantage in that it allows reducing the complex triangular XXZ model to a simpler honeycomb boson
model.
However, since $V$ becomes large when $J_{xy}/J_{zz} \ll 1/S$, the strong anisotropy also drives
a breakdown of spin-wave theory, making our results inapplicable when $J_{xy}$ is too small.
To access the large-$V$ limit, it would be necessary to work within a projected Hilbert space.
Alternatively, it is not excluded that one may find a coupling constant renormalization, such that
the effective coupling $V_{\rm R}$ remains finite at $V \to \infty$.
(For example, in the low-density expansion in superfluids strong short-range interactions are
renormalized by trading bare couplings for a $t$-matrix~\cite{beliaev_jetp_1958b, fisher_prb_1988}).
Here, we do not attempt to develop a renormalization of $V$, and confine ourself to
studying the first corrections in a ``bare'' perturbation theory in $V$ (although we do
include renormalizations of the gap, which are essential even in the small $V$ region to
regularize infrared divergences).

This means that the results of this work cannot give conclusive answers concerning the spectrum of
KCSO, which eventually, requires an analysis of the large-$V$ limit.
We can, nevertheless, use our results to give a first qualitative discussion in this direction.
In particular, we can discuss the qualitative structure of the corrections assuming small $V$, in
comparison with KCSO.
This is in part motivated by the fact that linear spin-wave theory, while missing the presence of
roton minima, gives predictions which are surprisingly close to experiments on other aspects of the
spectrum.
First, LSWT predicts two almost degenerate branches of excitations at zero fields, which split at
finite field (a prediction, which qualitatively is consistent with measurements reported in
Ref.~\cite{zhu_npj_2025}).
Secondly, the experiments indicate a small gap at the $K$ point.
If this mode is connected to the accidental degeneracy of the classical model, the smallness of
the gap, can be viewed as another suprisingly accurate prediction of spin-wave theory.
This motivates, as a first qualitative analyis, a discussion assuming small corrections to
spin waves.

From our analysis we find that describing the spectrum encounters difficulties, even in this
framework.
The pseudo-Goldstone gap grows rapidly with $V$ and reaches values comparable with experiments
already at $V \simeq 0.05$.
For such small values of $V$, on the other hand, there are only weak corrections at finite
$\bk$, and no roton minimum is generated.
In addition, although corrections produce a downward renormalization overall, the leading
corrections for $V$ very small actually go in the direction of producing a local maximum at $M$.
For larger $V$, the pseudo-Goldstone gap grows to too large values, before we can see an onset of a
roton minimum.

While the analysis which we present does not provide, directly an interpretation of the spectrum of
KCSO, it is justified in the case of a magnet with very large $S$.
In this framework, our analysis provides a discussion of the impact of pseudo-Goldstone modes on
renormalizations at finite momenta.

This article is organized as follows.
In Sec.~\ref{model}, we introduce the model and we discuss the classical, the spin-wave, and the
perturbative limit $J_{xy} \ll 1$ for the XXZ model.
In this discussion, we follow some of the results presented in earlier studies.
In Sec.~\ref{NLSWT} we discuss a systematic reduction of the problem to a honeycomb-lattice
effective model, in the limit $S \to \infty$, $J_{xy} \to 0$ at constant $V = J_{zz}/(S J_{xy})$.
The effective model resulting from this limit consists of soft-core bosons with
nearest-neighbour repulsion, and is used throughout the following sections.
In Sec.~\ref{semiclassical} we analyze the model in the semiclassical limit.
We discuss, in particular, the degeneracy of the classical ground state for $b = 0$, which has
a particularly simple form, and which can be traced to a simple symmetry of the classical energy.
In Sec.~\ref{self_energy} we present analytical expressions for the self-energy at one-loop order.
These expressions are applied in Sec.~\ref{pseudo-Goldstone_gap} to study the gap of the
pseudo-Goldstone branch and in Sec.~\ref{corrections_to_the_spectrum} to study the corrections to
the spectrum at finite momenta.
Finally, we comment on the applicability of the results to KCSO, showing that the nonlinear
spin-wave analysis is inadequate to describe some crucial spectral features observed experimentally
in this compound.
In Sec.~\ref{summary_and_conclusions} we summarize and conclude the article.

\section{Model, linear spin-wave approximation, and perturbation theory}
\label{model}

The starting point of our analysis is the XXZ model in a longitudinal magnetic field, whose
Hamiltonian reads:
\begin{equation} \label{XXZ}
\begin{split}
H & = \sum_{\langle i, j \rangle} \left( J_{zz} S^{z}_{i} S^{z}_{j} + J_{xy}
\left(S^{x}_{i} S^{x}_{j} + S^{y}_{i} S^{y}_{j}\right)\right) \\
& - h  \sum_{i} S^{z}_{i}~.
\end{split}
\end{equation}

We focus in particular on the case of a strong antiferromagnetic exchange $J_{zz} > 0$, such
that the ratio $\alpha = J_{xy}/J_{zz} \ll 1$ is small, and we restrict our analysis to the
low-field phase.

Within the classical and the linear spin-wave approximation, the model has been discussed
intensively~\cite{miyashita_jpsj_1985, miyashita_jpsj_1986, kleine_zpb_1992, sheng_jpcm_1992,
murthy_prb_1997}.
In this section, we briefly discuss some properties of these approximations, summarizing also some
results of earlier studies.

The ground state in the classical approximation can be easily shown to be ordered, with a
three-sublattice unit cell.
It can be shown in addition that the average magnetic moments on the three sublattices, $\bS_{\rm
A}$, $\bS_{\rm B}$, and $\bS_{\rm C}$, are coplanar, and belong to a common plane containing the
$z$ axis~\cite{miyashita_jpsj_1985, miyashita_jpsj_1986}.

At strictly zero field, however, the classical energy does not fix a unique solution for the ground
state.
The system possesses instead an accidental degeneracy.
There is a one-parameter family of inequivalent ground states, with different moments $\bS_{\rm A}$,
$\bS_{\rm B}$, $\bS_{\rm C}$, on the three sublattices, all having the same energy $E_{\rm cl} = - N
(1 + \alpha + \alpha^{2}) S^{2} J_{zz}/(1 + \alpha)$~\cite{fazekas_pm_1974, miyashita_jpsj_1985,
kleine_zpb_1992}.
In particular, it can be shown that it is possible to find a ground state for any choice of the
classical moment on one of the three sublattices.
In other words, if we choose $\bS_{\rm C}$ to have an arbitrary orientation, we can always adjust
$\bS_{\rm A}$ and $\bS_{\rm B}$ as a function of the direction of $\bS_{\rm C}$ in such way that
the configuration $(\bS_{\rm A}, \bS_{\rm B}, \bS_{\rm C})$ is a classical ground state.

This degeneracy is simple to understand in the strict limit $J_{xy} = 0$, when the Hamiltonian
reduces to the Ising term $H = \sum_{\langle i, j \rangle} J_{zz} S^{z}_{i} S^{z}_{j}$.
In this limit, the classical ground states are all configurations satisfying a simple rule: that
every triangular plaquette in the lattice contains at least a pair of equal and opposite spins
collinear with the $z$ axis (with $\theta= 0$ and $\theta = \pi$ respectively).
This rule is satisfied by a macroscopic number of non-periodic configurations, which can be derived
from the ground states of the Ising model~\cite{wannier_pr_1950} by promoting the spins to
classical vectors.

However, the ground-state still possesses a degeneracy if we consider only periodic configurations
with three-sublattice order.
For example we can choose $\theta_{\rm A} = 0$ on sublattice A, $\theta_{\rm B} = \pi$ on sublattice
B, and an arbitrarily oriented magnetization on sublattice C.
Thus, in the Ising case, one can find a ground state for an arbitrary choice of $\bS_{\rm C}$.

When a nonzero $J_{xy}$ us turned on, the solutions with three-sublattice periodicity become
stabilized classically, as they have lower classical energy than non-periodic configurations.
However crucially, the degeneracy among different ordered states with three
sublattice unit cell, persists classically, even at finite $J_{xy}$~\cite{miyashita_jpsj_1985,
kleine_zpb_1992}.

\begin{figure}[t]
\centering
\includegraphics[scale=1]{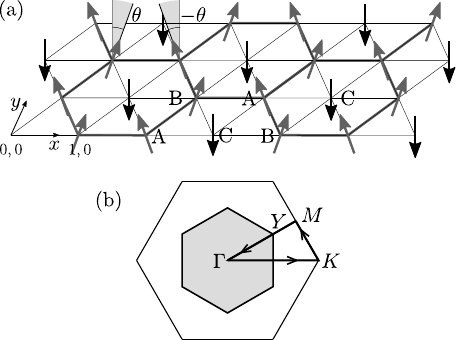}
\caption{\label{3subl} (a) Structure of the ``Y'' configuration. (b) Brillouin zone and special
$\bk$ points in momentum space.
The coordinates of the $\Gamma$, $K$, $M$, and $Y$ points are respectively $\Gamma = (0, 0)$, $K =
(4\pi/3, 0)$, $M = \pi(1, \sqrt{3}/3)$, $Y = 2 \pi/3 (1, \sqrt{3}/3)$.
The gray area shows the reduced Brillouin zone, folded due to the $\sqrt{3} \times \sqrt{3}$ unit
cell.
}
\end{figure}

The selection of a single ground state eventually occurs due to corrections beyond the classical
limit (i.e. by a quantum order-by-disorder effect)~\cite{kleine_zpb_1992, kleine_zpb_1992b,
murthy_prb_1997}.
By a linear-spin-wave analysis, Ref.~\cite{kleine_zpb_1992} showed that the state with the lowest
energy is the ``Y'' configuration, shown in Fig.~\ref{3subl}.
In this configuration, one of the sublattices is collinear and directed along the negative $z$
direction, while the spins on the remaining two sublattices are canted, and form equal and opposite
angles with respect to the $z$ axis. (The three moments are thus oriented like the three segments
of the letter ``Y'').
The spin moments are $\bS_{\rm A} = (\sin \theta, 0, \cos \theta)$, $\bS_{\rm B} = (-\sin \theta,
0, \cos \theta)$, $\bS_{\rm C} = (0, 0, -1)$ with $\cos \theta = 1/(1 + \alpha)$.

For nonzero magnetic field the degeneracy of the ground state is instead lifted already at
the classical level.
It can be shown that a small field stabilizes the same Y-shaped configuration which is selected by
the spin-wave fluctuations~\footnote{See Ref.~\cite{murthy_prb_1997} for a related analysis in the
case $J_{xy} < 0$.}.
Thus, in the semiclassical approximation, one can consistently assume a Y state in the low field
region, both at zero and at small finite field.
As the field grows, the canting angle $\theta$ decreases and the spins become more aligned in the
direction of the $z$ axis.
By a simple minimization of the classical energy, it can be seen that the canting angle $\theta$
depends on the field according to the relation $\cos \theta = (1 + \alpha b/3)/(1 + \alpha)$, where
$b = h/(S J_{xy})$.
At $b = 3$, the canting closes, and the spins in the A and B sublattice become collinear.
This marks the transition to a gapped $1/3$-magnetization plateau.
In this work, we restrict our analysis to the low-field canted state in the region $0 < b < 3$, and
we do not discuss other phases of the model.

The excitation spectrum implied by the semiclassical theory consists of spin-wave excitations which
in the large-$S$ limit can be described by linear-spin wave theory.
Due to the three-sublattice periodicity, the spectrum consists of three magnon branches, and in
general has to be determined by solving a three-sublattice problem~\cite{kleine_zpb_1992,
jia_prr_2024}.

However, in the limit of small $\alpha = J_{xy}/J_{zz}$, Ref.~\cite{kleine_zpb_1992} showed that
the problem can be significantly simplified.
In fact, for $\alpha \ll 1$ the canting angle $\theta$ is small.
As a result, the C spins are surrounded by 6 spins which are almost collinear to the positive $z$
axis.
This results in a strong Ising exchange field on the C spins, which pushes the corresponding spin
waves to high energy.
As a result, the C spins can be considered as ``frozen'' in the linear spin wave approximation for
$\alpha$ small~\cite{kleine_zpb_1992}.

Using this simplification, Ref.~\cite{kleine_zpb_1992} showed that the spectrum of the low energy
branches can be derived analytically, using an effective model on the honeycomb lattice (composed
by the A and B sites).
The dispersion of the two low-energy magnons for an arbitrary field $0 < b < 3$ and for
$J_{xy} \ll J_{zz}$ is~\cite{kleine_zpb_1992}:
\begin{equation}\label{LSWT_ap}
\epsilon_{\lambda}(\bk) = S J_{xy} \sqrt{9 - \lt(1 - \frac{2}{3}b \rt)|f_{\bk}|^{2} + 2 \lambda b
|f_{\bk}|}~.
\end{equation}
Here $\lambda = \pm 1$ is a branch index identifying the two magnon modes and $f_{\bk} = {\rm
e}^{ik_{x}} + {\rm e}^{-i(k_{x} + \sqrt{3} k_{y})/2} + {\rm e}^{-i(k_{x} - \sqrt{3} k_{y})/2}$.
The $\lambda = -1$ mode is gapless for $\bk \to 0$ and represents the Goldstone mode
associated with the broken $U(1)$ symmetry.
The $\lambda = +1$ mode instead is gapped for $b > 0$ and becomes gapless, in the
linear spin wave approximation, for $b = 0$.
It corresponds to a pseudo-Goldstone mode~\cite{murthy_prb_1997}, reflecting the degeneracy of the
$b = 0$ ground state.

As a remark, we note that the dispersion relations~\eqref{LSWT_ap} are identical in
form to those studied in Ref.~\cite{maksimov_prb_2016}, which analyzed the easy-plane XXZ
model in a longitudinal field on the honeycomb lattice.

\begin{figure}[t]
\centering
\includegraphics[scale=1]{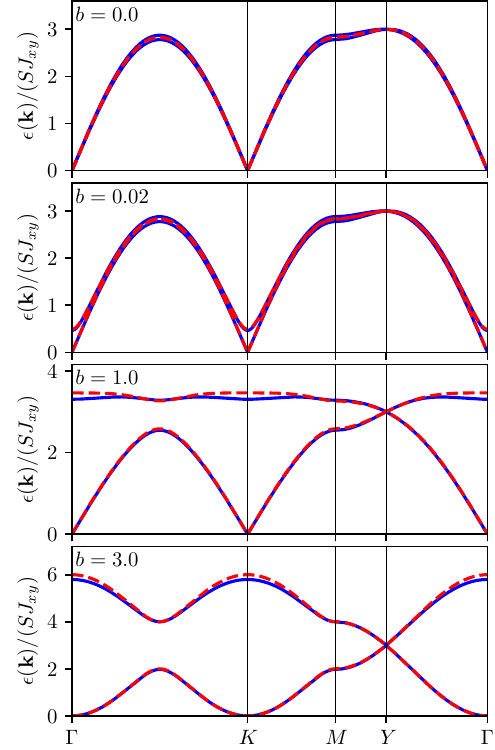}
\caption{\label{LSWT} Dispersions of the two lowest LSWT branches, calculated for $\alpha = 0.07$
and for magnetic fields $b = 0$, $b = 0.02$, $b = 1$, and $b = 3$, along the path $\Gamma K M
\Gamma$ shown in Fig.~\ref{3subl}.
The blue solid lines show the dispersions obtained by solving the complete LSWT equations for the
triangular XXZ model; the red dashed lines show the analytical dispersions~\eqref{LSWT_ap}, which
are the $\alpha \to 0$ limits of the LSWT.
Due to the small value of $\alpha$, the dispersions are well approximated by the analytical
expressions.
The dispersions at zero field present saddle points at $M = (\pi, \sqrt{3}\pi/3 )$.
In addition, the two dispersions cross at the point $Y = (2 \pi/3, 2\sqrt{3}\pi/9)$, where
$f_{\bk} = 0$ for all values of $b$.
The Y point is akin to a ``Dirac'' node.
Due to this reason, the separation of the spectrum into a Goldstone and a pseudo-Goldstone part is
only meaningful at small momentum, and cannot be continuously extended to the full Brillouin zone.}
\end{figure}

The LSWT dispersions~\eqref{LSWT_ap} along high-symmetry directions are shown in Fig.~\ref{LSWT} in
dashed red lines.
From Fig.~\ref{LSWT}, it is clear that the LSWT spectrum completely misses the presence of a roton
minimum, as that observed experimentally in KCSO~\cite{zhu_npj_2025, zhu_npj_2025}.
The LSWT approximation, in fact, does not predict a local minimum at M, but rather a saddle point
(with a minimum along $\Gamma-M$ and a maximum along $K-M$ directions).

In Fig.~\ref{LSWT} we also represented the spectra of the two lowest energy modes using a full
linear spin-wave calculation, solving the full three-band magnon problem, assuming an anisotropy
ratio $\alpha = 0.07$.
As it can be seen in the figure, the full dispersions are hardly distinguishable from the
analytical expressions~\eqref{LSWT}, and similarly, do not present any trace of a roton like
minimum.
Thus, as explained in Ref.~\cite{zhu_npj_2025, zhu_prl_2024}, the linear-spin-wave approximation
does not provide a satifsfactory explanation of the experimental observations.
The roton-like minimum thus cannot be explained in terms of free magnons but requires an
understanding in terms of interactions (or, alternatively, in terms of a spinon-picture as proposed
in Ref.~\cite{bose_prb_2025}).

While the $\alpha \to 0$ limit brings a simplification at the level of linear spin wave theory, it
was noted already in Ref.~\cite{kleine_zpb_1992b} that this limit is, in fact, problematic in the
spin-wave expansion.
In fact, if $\alpha \to 0$ for fixed $S$, the Ising energy $J_{zz} S^{z}_{i} S^{z}_{j}$ becomes
dominant and acts effectively as a constraint.
As a result, the low-energy states for $\alpha \to 0$ are governed by an effective Hamiltonian
$H_{\rm P} = \sum_{\langle i, j \rangle} J_{xy} P \left(S^{x}_{i} S^{x}_{j} + S^{y}_{i}
S^{y}_{j}\right) P - h \sum_{i} P S^{z}_{i} P$, where $P$ is the projector on the
subspace of ground states of the Ising energy~\cite{kleine_zpb_1992, moessner_prb_2001,
fazekas_pm_1974, sen_prl_2008, wang_prl_2009, zhu_npj_2025}.
For $S  = 1/2$, this projected Hamiltonian is the well known ``quantum Ising
model''~\cite{moessner_prb_2001}, and is equivalent to a quantum dimer model on the honeycomb
lattice~\cite{fazekas_pm_1974, sen_prl_2008, wang_prl_2009, zhu_npj_2025}.
However, also for any, arbitrarily large value of $S$, the projected model is a strongly coupled
many-body problem without any small expansion parameter~\cite{kleine_zpb_1992b}.

\begin{figure}[t]
\centering
\includegraphics[scale=1]{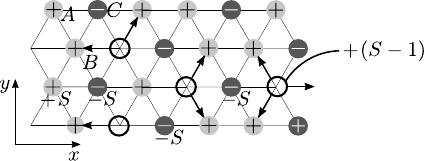}
\caption{\label{strong_coupling_3subl} The ground state for $\alpha \to 0$ and zero field
hypothesized by Ref.~\cite{kleine_zpb_1992b}.
For $S\geq 3/2$, the spins at the C sublattice are frozen in the state $S^{z} = -S$.
The remaining spins fluctuate between $S^{z} = +S$ and $S^{z} = +(S-1)$ and form a superfluid of
bosons hopping on the honeycomb lattice.
In the figure, the $-S$ spins are represented by black circles, the $+S$ spins by gray circles, and
the $+(S -1)$ spins by open circles.
The Ising constraint for $\alpha \to 0$ is equivalent to an infinite nearest-neighbour repulsion
between the $S^{z} = +(S-1)$ spins.}
\end{figure}

Ref.~\cite{kleine_zpb_1992b} presented qualitative arguments indicating that the ground state, even
in the strong-coupling regime, should have the same structure as the semiclassical state at zero
field, at least for $S \geq 3/2$.
In particular, the analysis of Ref.~\cite{kleine_zpb_1992b} indicated a ground state in which the
spins at C sites are frozen in the $S^{z} = -S$ state, whereas the spins in the A and B
sublattices form a superfluid, as in the semiclassical Y state (see
Fig.~\ref{strong_coupling_3subl}).

For $S = 1/2$, the exact diagonalization analyses presented in Ref.~\cite{ulaga_prb_2025}
indicate the persistence of the same pattern of lattice symmetry breaking.
However, we note that the results of Ref.~\cite{ulaga_prb_2025, ulaga_arxiv_2025} indicate a
``solid'', and not a ``supersolid'' state in the $S = 1/2$ case, and for zero total magnetization,
which contrasts with semiclassical predictions.
In the case of an effective honeycomb XXZ model with nearest-neighbour interactions, instead,
Ref.~\cite{ulaga_arxiv_2025} indicated the persistence of a finite superfluid density.

\section{Effective model on the honeycomb lattice}
\label{NLSWT}

As summarized in the earlier section, the analyses of Kleine~\emph{et al.}~\cite{kleine_zpb_1992,
kleine_zpb_1992b} indicate for a fixed $S \geq 3/2$, a spin-supersolid ground state, with a Y order,
which evolves between a semiclassical behavior at $V = 1/(S \alpha) \ll 1$ to a strong-coupling
limit at $V \gg 1$.
For $S = 1$ and $S = 1/2$ the problem is in general more complex~(see Refs.~\cite{fazekas_pm_1974,
kleine_zpb_1992b, burkov_prb_2005, sen_prl_2008, wang_prl_2009, heidarian_prl_2010, zhang_prb_2011,
ulaga_prb_2025, zhu_npj_2025, gallegos_prl_2025} for the $S = 1/2$ case).
However, since in this work we study the semiclassical expansion, our analysis is insensitive to a
change of behavior at finite $S$.

Within the spin-wave expansion, we assume therefore that the qualitative picture of the Y state is
valid.
In this section, we use these considerations and derive a simple effective model, which interpolates
between the $V \ll 1$ and the $V \gg 1$ regimes, while at the same time, taking advantage of
simplifications due to the smallness of the ratio $\alpha = J_{xy}/J_{zz}$.

The crossover at $\alpha \simeq 1/S$ implies that the limits $S \to \infty$ and $\alpha \to 0$ do
not commute.
Thus, the limit $\alpha \to 0$ cannot be taken order by order in the spin-wave expansion.
The fact that the parameter $V = 1/(S \alpha)$ plays a crucial role, however, suggests to take the
limit by sending $S \to \infty$ and $\alpha \to 0$ keeping the combination $V$ fixed.

This limit, does not present any difficulty because the parameter $V$ which controls the coupling
strength~\cite{kleine_zpb_1992b} is held finite.
However, it is simple to see that sending $S \to \infty$, $\alpha \to 0$ at $V$ fixed leads to a
simplification of the model.
In fact, in this limit, it is justified to replace $S^{z}_{i} = -S$ at all C sites (by the
considerations summarized in Sec.~\ref{model}).
Thus we can consider, instead of the triangular XXZ model, an effective
XXZ model on the honeycomb lattice, in which the effect of the C sites is replaced by that of a
static field.
In addition, a further simplification can be applied, after introducing a Holstein-Primakoff
representation of the spin operators:
\begin{equation} \label{HP}
\begin{split}
S^{x}_{i} & = \frac{1}{2} \lt(S^{+}_{i} + S^{-}_{i}\rt)~, \qquad S^{y}_{i}  = \frac{1}{2i}
\lt(S^{+}_{i} - S^{-}_{i}\rt)~, \\
S^{+}_{i} & = \lt(\sqrt{2 S - n_{i}}\rt) a_{i}~, \qquad S^{-}_{i}  = a^{+}_{i} \sqrt{2 S -
n_{i}}~,\\
S^{z}_{i} & = S - n_{i}~, \qquad n_{i} = a^{+}_{i} a_{i}~.
\end{split}
\end{equation}

At the classical level (for $S \to \infty$), the average density of bosons can be determined by
identifying $\langle S^{z}_{i} \rangle_{0} = S - \langle a^{+}_{i} a_{i}\rangle_{0} = S - \langle
n_{i} \rangle_{0} = S \cos \theta$, where $\theta$ is the classical canting angle.
(Here $\langle .. \rangle_{0}$ stands for an average in the classical state).
Using $\cos \theta = (1 + \alpha b/3)/(1 + \alpha) = (1 + b/(3 S V))/(1 + 1/(SV))$ and taking $S\to
\infty$ we see that for $V$ fixed:
\begin{equation}
\langle n_{i} \rangle_{0} \simeq  S (1 - \cos \theta) = (1 - b/3)/V~.
\end{equation}

This implies that the number of bosons is of order 1, and not of order $S$, even if the state is
canted (the canting angle reduces together with $S$ in such way that $n$ remains finite).
This allows one to truncate Eq.~\eqref{HP} for $a^{+}_{i} a_{i} \ll S$, replacing $S^{+}_{i} \to
\sqrt{2S} a_{i}$, $S^{-}_{i} \to \sqrt{2S} a^{+}_{i}$.
Using this truncation, we obtain an effective Hamiltonian on the honeycomb lattice:
\begin{equation} \label{H_ef_0}
\begin{split}
H &  = - N S^{2} J_{zz} - \frac{1}{3} N S h + S J_{xy} \sum_{\langle i, j \rangle} \lt(a^{+}_{i}
a_{j} + a^{+}_{j} a_{i}\rt) \\
& \qquad +  J_{zz} \sum_{\langle i, j \rangle} n_{i} n_{j} + h \sum_{i} n_{i}~.
\end{split}
\end{equation}

For simplicity in the following, we absorb the constant $- N S^{2} J_{zz} -  N S h/3$ in a shift of
the energy scale, and we use units in which $S J_{xy} = 1$ and $\hbar=1$.
The effective Hamiltonian on the honeycomb lattice,  then becomes:
\begin{equation} \label{H_ef}
H = \sum_{\langle i, j \rangle} \lt(a^{+}_{i} a_{j} + a^{+}_{j} a_{i}  + V n_{i} n_{j} \rt) + b
\sum_{i} n_{i}~.
\end{equation}

The model thus reduces to a soft-core boson model with nearest-neighbour repulsion $V = 1/(S
\alpha)$.
As $V$ ranges from small to large values, the model interpolates between a semiclassical region,
which can be described by the loop expansion, and the strong-coupling regime discussed in
Ref.~\cite{kleine_zpb_1992b}.

We note that similar descriptions, in terms of effective bosonic models on the honeycomb lattice
have been used to describe the triangular XXZ model in Ref.~\cite{zhang_prb_2011, ulaga_prb_2025}.
However, in these works, the mapping which was considered involved hard-core bosons instead of
soft-core ones.
In the limit $V \to \infty$ it can be expected that the system fluctuates only among configurations
with $n_{i} = 0, 1$ thus recovering the hard-core limit~\cite{kleine_zpb_1992b}.

We note in addition that the reduction to a honeycomb model presented in Ref.~\cite{zhang_prb_2011}
is only approximate for $S = 1/2$ because in this case the C sites can be flipped without leaving
the ground state manifold.
The derivation presented here, instead, in analogy with the discussion of
Ref.~\cite{kleine_zpb_1992}, leads to an exact reduction of the model in the limit $S \to \infty$,
$\alpha \to 0$.

As a final remark, we note that the approximation of frezing the C sublattice is valid only when
$\alpha \ll 1$.
We expect instead important quantitative differences between the honeycomb model and the full
three-band model when $\alpha$ is larger.
In particular, we expect significant quantitative differences between the nonlinear effects studied
in the next sections, and the $1/S$ corrections of the triangular XXZ model for parameters relevant
to the spin-1/2 compound Na$_{2}$BaCo(PO$_{4}$)$_{2}$~\cite{gao_prb_2024, jia_prr_2024,
gao_npj_2022, chi_prb_2024} and to the recently analyzed spin-5/2 triangular magnet
K$_{2}$Mn(SeO$_{3}$)$_{2}$~\cite{zhu_arxiv_2026}.

\subsection*{Cartesian representation}

In most of the following calculations, we find it convenient to work in cartesian coordinates,
introducing the operators $m^{\alpha}_{i} = (m^{x}_{i}, m^{y}_{i})$, $m^{x}_{i} = (a_{i} +
a^{+}_{i})/\sqrt{2}$, $m^{y}_{i} = -i (a_{i} - a^{+}_{i})/\sqrt{2}$.
Identifying $m^{1}_{i} = m^{x}_{i}$ and $m^{2}_{i} = m^{y}_{i}$, the commutation
relations can be written as $[m^{\alpha}_{i}, m^{\beta}_{j}] = -\sigma_{y}^{\alpha \beta}
\delta_{ij}$.
The Hamiltonian~\eqref{H_ef} can be represented as
\begin{equation}\label{H_ef_1}
\begin{split}
H & = \sum_{\langle i, j \rangle} \big(m^{\alpha}_{i} m^{\alpha}_{j} + \frac{V}{4}
M^{\alpha \beta} M^{\gamma \delta} m^{\alpha}_{i} m^{\beta}_{i} m^{\gamma}_{j}
m^{\delta}_{j}\big)\\
& \qquad + \frac{b}{2} \sum_{i} M^{\alpha \beta} m^{\alpha}_{i} m^{\beta}_{i}~,
\end{split}
\end{equation}
where the matrix $M$ is defined as $M^{\alpha \beta} = \delta^{\alpha \beta} - \sigma_{y}^{\alpha
\beta}$.
The term $-\sigma_{y}^{\alpha \beta}$ accounts for the ordering of creation and annihilation
operators ($n_{i} = M^{\alpha \beta} m^{\alpha}_{i} m^{\beta}_{i}/2$, $n_{i} + 1 = M^{\alpha \beta}
m^{\beta}_{i} m^{\alpha}_{i}/2$).

\section{Semiclassical limit $V \to 0$}
\label{semiclassical}

Expanding in powers of $V$ generates the semiclassical expansion, which diagrammatically
corresponds to the loop expansion~\footnote{In a path integral representation, after the change of
variables $m \to m/\sqrt{V}$, the action associated with the Hamiltonian~\eqref{H_ef_1} becomes
multiplied by an overall factor $1/V$; thus the expansion in powers of $V$ is equivalent to the loop
expansion in analogy with other field theories~\cite{zinn-justin_qft}.}.
The leading order $V \to 0$ is the classical limit of the model and mirrors the classical limit
discussed in Sec.~\ref{model}.
However, some results, for example the presence of an accidental classical degeneracy, take a
simpler form than in the full XXZ model.
Thus, in this section, we find it useful to rederive the classical and the spin-wave
approximations, directly for the effective model~\eqref{H_ef_1}.

At leading order the Bose operators $a_{i}$ can be replaced with complex amplitudes $\varphi_{i}$,
and the order parameters are found by minimizing the classical energy
\begin{equation}\label{E_cl}
\begin{split}
E(\varphi_{i}) & = \sum_{\langle i, j \rangle} \lt(\varphi_{i}^{*} \varphi_{j} +
\varphi_{j}^{*}
\varphi_{i} + V |\varphi_{i}|^{2} |\varphi_{j} |^{2} \rt) \\
& + b \sum_{i} |\varphi_{i}|^{2}~.
\end{split}
\end{equation}

Assuming a two-sublattice ordering with $\varphi_{i} = \varphi_{\rm A}, \varphi_{\rm B}$ on
sublattices A and B, respectively, the energy is $E = N \lt(\varphi_{\rm A}^{*} \varphi_{\rm B} +
\varphi_{\rm B}^{*} \varphi_{\rm A} + V |\varphi_{\rm A}|^{2} |\varphi_{\rm B}|^{2}\rt) +
\frac{1}{3} N b \lt(|\varphi_{\rm A}|^{2} + |\varphi_{\rm B}|^{2}\rt)$.
Here $N$ is defined as the total number of triangular lattice sites, including the ``frozen'' C
sites; thus the honeycomb lattice model has $2N/3$ sites.

For $0 < b < 3$ and $b \neq 0$, the energy minimum is unique up to $U(1)$ rotations, and has equal
and opposite amplitudes on the two sublattices: $\varphi_{\rm A} = {\rm e}^{i \phi} \sqrt{(1 -
b/3)/V} $, $\varphi_{\rm B} = - \varphi_{\rm A}$.
At zero field, however, the ground state shows an additional accidental degeneracy, which
is a remnant of the degeneracy of the XXZ model.
The classical energy for $b = 0$ depends only on the product $\rho = \varphi_{\rm B}^{*}
\varphi_{\rm A}$.
Any pair of $\varphi_{\rm A}$, $\varphi_{\rm B}$ for which $\rho$ has the optimal value $\rho=-1/V$
is a ground state.
We checked that the quantum corrections select as ground state the Y configuration with
$\varphi_{i} = r_{i}/\sqrt{V}$, $r_{i} = \pm 1$ on sublattices A and B, consistently with earlier
analyses~\cite{kleine_zpb_1992}.

As a remark, we note that the degeneracy can be expressed in terms of a symmetry of the
classical energy~\eqref{E_cl} for $b = 0$: the symmetry under rescaling $\varphi_{i} \to \xi_{i}
\varphi_{i}$, with $\xi_{i} = \lambda, 1/\lambda$ on sublattices A and B, respectively.
Interestingly this transformation leaves $E$ invariant not only for configurations of $\varphi_{i}$
which minimize the energy, but for an arbitrary configuration of the complex amplitudes
$\varphi_{i}$.
The rescaling however cannot be extended to a symmetry at the quantum level, as the
rescaling violates the commutation relations $[a_{i}, a^{+}_{i}] \to \xi_{i}^{2} \neq 1$.
This implies that the degeneracy is lifted by quantum fluctuations, leading, as in the
XXZ model, to ground state selection, and a gap to the pseudo-Goldstone mode.
In a path integral formulation, the ``staggered rescaling'' transformation leaves the potential
invariant, but changes the kinetic term in the action.
This fact suggests the possibility to derive a Ward identity.
We leave this for future study.

\subsection*{Linear spin wave spectrum}
\label{propagator}

The magnon spectrum in the semiclassical limit can be determined by expanding Eq.~\eqref{H_ef} near
the classical minimum and by keeping terms of quadratic order in the
fluctuations.
By solving the corresponding equations of motion, we find the bare magnon
dispersions~\cite{kleine_zpb_1992}
\begin{equation} \label{LSWT_ap_res}
\epsilon_\lambda(\bk) = \sqrt{9 - \lt(1 - \frac{2}{3} b\rt)|f_{\bk}|^{2} + 2\lambda b |f_{\bk}|}~,
\end{equation}
which are identical to Eq.~\eqref{LSWT_ap} in rescaled units.
As explained in Sec.~\ref{model}, the modes $\lambda = \pm 1$ correspond at long-wavelengths to
Goldstone and pseudo-Goldstone modes.

\section{Self-energy in the one-loop approximation}
\label{self_energy}

To calculate the corrections to the spectrum we compute the self-energy at one-loop order. The
spectrum is then determined by the Dyson equation $G^{-1} = G_{0}^{-1} - \Sigma$.
The Green function and the self-energy are matrices carrying two cartesian indices.
The matrices contain (in a different basis) both the normal and the anomalous self-energy terms in
the Beliaev-Dyson equation~\cite{beliaev_jetp_1958b, castellani_prl_1997} (see also
Ref.~\cite{auerbach_magnetism}).

\subsection{Bare propagator}

As a first step, we calculate the bare propagator in the cartesian coordinate representation,
defined as: $G_{0ij}^{\alpha \beta}(t - t') = -i \langle 0 |T\{\delta m^{\alpha}_{i}(t) \delta
m^{\beta}_{j}(t')\} |0 \rangle$, with $\delta m = m - \langle m \rangle$.

Using the linear-spin wave equations of motion (see Appendix~\ref{bare_propagators}) we find in
Fourier space
\begin{equation}\label{G0}
\begin{split}
G^{\alpha \beta}_{0 ij}(t-t') & = \int_{-\infty}^{\infty} \frac{{\rm d}\omega}{2\pi} \int_{{\rm BZ}}
\frac{{\rm d}^{2}k}{\Omega} \\
& G_{0\tau_{i} \tau_{j}}^{\alpha \beta}(\bk, \omega)  {\rm e}^{i \bk \cdot (\bx_{i} - \bx_{j}) - i
\omega(t-t')}~,
\end{split}
\end{equation}
with
\begin{equation}\label{G_unitary}
\begin{split}
G^{\alpha \beta}_{0 \tau \tau'}(\bk, \omega) = \sum_{\lambda = \pm 1} U_{\tau
\lambda}(\bk) G_{0\lambda}^{\alpha \beta}(\bk, \omega)U^{*}_{\tau' \lambda}(\bk)~,
\end{split}
\end{equation}
and
\begin{equation} \label{G0_propagator}
\begin{split}
& G^{\alpha \beta}_{0 \lambda}(\bk, \omega)  = \frac{1}{\omega^{2} -
\epsilon_{\lambda}^{2}(\bk) + i0^{+}} \Bigg[- \omega \sigma^{\alpha \beta}_{y} \\
&  + \lt(3 +
\frac{b}{3} \lambda |f_{\bk}|\rt) \delta^{\alpha \beta}  + \lt(1 - \frac{b}{3}\rt)
\lambda |f_{\bk}| \sigma_{z}^{\alpha \beta}\Bigg] ~.
\end{split}
\end{equation}

In these expressions, $\int_{{\rm BZ}}$ is an integral over the Brillouin zone (BZ) (the shaded area
in Fig.~\ref{3subl}) and $\Omega = 8 \pi^{2} \sqrt{3}/9$ is the Brillouin-zone area.
The lower indices $\tau$, $\tau'$ are sublattice indices ($\tau=1$ for sublattice A and $\tau=2$
for sublattice B).
The indices $\tau_{i}$, $\tau_{j}$ in Eq.~\eqref{G0} denote the sublattices to which the sites $i$,
$j$ belong.
Finally, the upper Greek indices denote the cartesian coordinates ($\alpha = 1$ for the $x$
components, parallel to the average moments, and $\alpha = 2$ for the transverse, $y$ components),
and the index $\lambda$ labels the magnon branches ($\lambda = \pm 1$).
The bare propagator is diagonal in the branch index as shown by Eq.~\eqref{G_unitary}.
The transformation from the branch to the sublattice basis is defined by the unitary transformation
\begin{equation} \label{U_mat}
\begin{split}
U(\bk)  = \begin{vmatrix}
U_{11}(\bk) & U_{1 -1}(\bk) \\
U_{2 1}(\bk) & U_{2 -1}(\bk)
\end{vmatrix} & \\
 = \frac{1}{\sqrt{2 |f_{\bk}|}}  \begin{vmatrix}
\sqrt{f^{*}_{\bk}} & \sqrt{f^{*}_{\bk}} \\
\sqrt{f_{\bk}} & - \sqrt{f_{\bk}}
\end{vmatrix}
\end{split}
\end{equation}

To fix the signs in the definition of $U(\bk)$ we consider $\sqrt{f}$ to be everywhere
the complex square root of $f$ with positive real part.

The Green function propagates two magnons with the bare dispersions $\epsilon_{\lambda}(\bk)$,
$\lambda = \pm 1$.
The residue at the poles $\omega = \pm \epsilon_{\lambda}(\bk)$ can be expressed introducing
\begin{equation}
N_{\lambda}^{\alpha \beta}(\bk) = \lt(3 + \frac{b}{3} \lambda |f_{\bk}|\rt) \delta^{\alpha
\beta}  + \lt(1 - \frac{b}{3}\rt) \lambda |f_{\bk}| \sigma_{z}^{\alpha \beta}~.
\end{equation}
and
\begin{equation} \label{w_vectors}
w_{\lambda}^{\alpha}(\bk) = \begin{pmatrix}
                             i \sqrt{3 + \lambda |f_{\bk}|}\\
                             \sqrt{3 - \lt(1 - \frac{2}{3} b\rt) \lambda|f_{\bk}|}
                            \end{pmatrix}~.
\end{equation}

We then find that the residues $N \pm \epsilon \sigma^{y}$ can be expressed as
\begin{equation} \label{residues}
\begin{split}
& N^{\alpha \beta}_{\lambda}(\bk) - \epsilon_{\lambda}(\bk) \sigma_{y}^{\alpha \beta} =
w_{\lambda}^{\alpha}(\bk) w_{\lambda}^{*\beta}(\bk) \\
& N^{\alpha \beta}_{\lambda}(\bk) + \epsilon_{\lambda}(\bk) \sigma_{y}^{\alpha \beta} =
w_{\lambda}^{*\alpha}(\bk) w_{\lambda}^{\beta}(\bk)~.
\end{split}
\end{equation}

The coefficients $w_{\lambda}(\bk)$ are related to the matrix elements for emission of a
magnon in branch $\lambda$ (in the non-interacting magnon approximation):
\begin{equation}
\langle \lambda, \bk|\delta m^{+\alpha}_{\tau}(\bk)|0 \rangle = \frac{1}{\sqrt{2
\epsilon_{\lambda}(\bk)}} U^{*}_{\tau \lambda}(\bk) w^{*\alpha}_{\lambda}(\bk)~,
\end{equation}
where $\delta m_{\tau}^{\alpha}(\bk)$ is the fluctuation of the magnetization in Fourier
space.
This coefficient could be derived equivalently by performing a Bogoliubov transformation
of the linear spin- wave Hamiltonian.

As a remark, we note that the Green function has a particularly simple form, because the sublattice
space can be diagonalized independently of the cartesian space.
This is a special property of the model analyzed here, which can be traced to the short range of the
exchange interactions.
If the model is studied by a Bogoliubov transformation, this simplification is reflected in the fact
that the problem can be solved by performing first a unitary transformation, and secondly a
single-band paraunitary transformation (see Ref.~\cite{maksimov_prb_2016}).

\subsection{Self-energy at one-loop order}

The spectrum in presence of interactions is determined by the Dyson equation
\begin{equation}
G^{-1 \alpha \beta}_{\tau \tau'}(\bk, \omega) = G_{0 \tau \tau'}^{-1 \alpha \beta}(\bk, \omega) -
\Sigma^{\alpha \beta}_{\tau \tau'}(\bk, \omega)~.
\end{equation}

Since the bare propagator is diagonal in the $\lambda$ basis, we find it convenient to calculate
the self-energy in this representation.
The Dyson equation reads:
\begin{equation}
G^{-1 \alpha \beta}_{\lambda \lambda'}(\bk, \omega) = G_{0 \lambda}^{-1 \alpha \beta}(\bk,
\omega) \delta_{\lambda \lambda'} - \Sigma^{\alpha \beta}_{\lambda \lambda'}(\bk, \omega)~,
\end{equation}
with
\begin{equation}
\begin{split}
G_{0 \lambda}^{-1}(\bk, \omega)  = & -\lt(3 + \frac{b}{3} \lambda |f_{\bk}|\rt)
\delta^{\alpha \beta} - \omega \sigma^{\alpha \beta}_{y} \\
 & + \lt(1 - \frac{b}{3}\rt) \lambda |f_{\bk}| \sigma_{z}^{\alpha \beta}~.
\end{split}
\end{equation}

\begin{figure}[t]
 \includegraphics[scale=1]{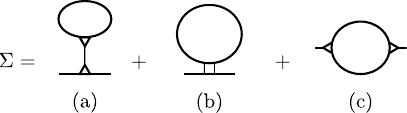}
 \caption{\label{diagrams} Self-energy diagrams contributing at one-loop order.
 The graphs (b) and (c) are one-line irreducible. The tadpole graph (a) instead is one-line
reducible.
The graph (a) reflects the change of the magnon Hamiltonian $\pa^{2} H/(\pa m_{i}^{\alpha} \pa
m_{j}^{\beta})|_{\bar{m}}$ due to the one-loop renormalization of the order parameter.
}
\end{figure}

At first order in $V$, the self-energy corrections are given by the one-loop diagrams in
Fig.~\ref{diagrams}.
The self-energy can be written as a sum $\Sigma = \Sigma_{{\rm a}} + \Sigma_{{\rm b}} +
\Sigma_{{\rm c}}$ of the three corresponding contributions.
As in other spin systems with canted ground states~\cite{chubukov_jpcm_1994, chernyshev_prb_2009,
maksimov_prb_2016b, rau_prl_2018}, the one-loop graphs involve a quartic vertex but also a cubic
vertex.

In our calculation, instead of computing all of the the diagrams directly, we find it convenient
to analyze the self-energy using the effective action approach~\cite{zinn-justin_qft}.
In this approach, the graphs~\ref{diagrams}b and~\ref{diagrams}c, which are one-particle
irreducible (1PI), give a direct contribution to the effective action.
The tadpole diagram~\ref{diagrams}a, instead, does not contribute because it is not one-particle
irreducible (1PI).
The effect of the diagram (a) however is encoded in the fact that, to calculate the Green
function, the effective action has to be expanded at its exact minimum.
This minimum is located at the exact value $\bar{m}_{i}^{\alpha}$ of the order parameter, and not at
the semiclassical minimum.
The self-energy $\Sigma_{{\rm a}}$ associated with the graph (a) can be calculated by taking into
account the shift $\bar{m}_{i}^{\alpha} - m_{i0}^{\alpha}$.

At one loop order, in particular, the self-energy term $\Sigma_{{\rm a}}$ can be calculated simply
by considering the variation of the non-interacting ``magnon Hamiltonian'' $\pa^{2} H/(\pa
m_{i}^{\alpha} \pa m_{j}^{\beta})|_{m_{i}^{\alpha} = \bar{m}_{i}^{\alpha}}$ due to the
correction of the order parameter $\bar{m}_{i}^{\alpha}$.
By an explicit calculation we find:
\begin{equation}\label{sigma_a}
\begin{split}
\Sigma_{{\rm a}\lambda \lambda'}^{\alpha \beta}(\bk, \omega) & = V(\varphi^{2} - \varphi^{2}_{\rm
cl})
\big((3 - \lambda |f_{\bk}|) \delta^{\alpha \beta} \\
& \qquad - \lambda |f_{\bk}|\sigma_{z}^{\alpha \beta}\big) \delta_{\lambda \lambda'}~,
\end{split}
\end{equation}
where $\varphi_{\rm cl}^{2} = (1 - b/3)/V$ is the square of the classical order parameter and
$\varphi^{2}$ is the square of the one-loop order parameter.
The correction $\varphi^{2} - \varphi_{\rm cl}^{2}$, in turn, can be calculated from a one-loop
tadpole diagram (see Appendix~\ref{order_parameter}).
We find:
\begin{equation}
\varphi^{2} - \varphi_{\rm cl}^{2} = c_{2} - c_{1}~,
\end{equation}
where $c_{1}$ and $c_{2}$ are static correlation functions calculated in the non-interacting
spin-wave model:
\begin{equation}\label{c1_c2}
\begin{split}
c_{1} & = \frac{1}{2} M^{\alpha \beta} \langle \delta m_{i}^{\alpha} \delta
m_{i}^{\alpha}\rangle_{0} ~, \\
& = - \frac{1}{2} + \sum_{\lambda = \pm 1} \int_{{\rm BZ}} \frac{{\rm d}^{2}k}{\Omega}
\frac{\lt(3 + \frac{b}{3} \lambda |f_{\bk}|\rt)}{4 \epsilon_{\lambda}(\bk)} \\
c_{2} & = \langle \delta m_{i}^{x} \delta m_{j}^{x} \rangle_{0}~\qquad ~( i,
j~\text{nearest-neighbour})\\
& = \frac{1}{12} \sum_{\lambda = \pm 1} \int_{{\rm BZ}} \frac{{\rm d}^{2}k}{\Omega} \frac{(3
+ \lambda |f_{\bk}|) \lambda |f_{\bk}|}{\epsilon_{\lambda}(\bk)}~.
\end{split}
\end{equation}

This completes the calculation of the diagram~\ref{diagrams}a.
The next graph,~\ref{diagrams}b, gives, also a frequency-independent self-energy.
The graph can be analyzed by representing the quartic interactions as
\begin{figure}[h]
\includegraphics[scale=1]{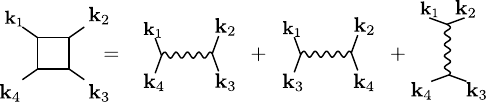}
\end{figure}

The self-energy term then consists of both Hartree- and Fock-type corrections.
We find that the self-energy $\Sigma_{{\rm b}}$ is (see Appendix~\ref{ap_self_energy}):
\begin{equation} \label{HF}
\begin{split}
\Sigma_{{\rm b} \lambda \lambda'}^{\alpha \beta}(\bk, \omega)  &=
\text{
\includegraphics[scale=1]{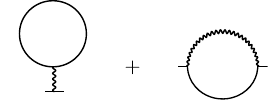}
}
\\
& ~ \\
 = V\Big(& 3 c_{1}  \delta^{\alpha \beta} + \frac{1}{2} \lambda |f_{\bk}|
\big((c_{2} + c_{3})\delta^{\alpha \beta} \\
& + (c_{2} - c_{3}) \sigma_{z}^{\alpha \beta} \big)\Big) \delta_{\lambda \lambda'}~,
\end{split}
\end{equation}
where $c_{1}$, $c_{2}$ are the coefficients in Eq.~\eqref{c1_c2} and $c_{3}$ is the
nearest-neighbour transverse correlation
\begin{equation}
\begin{split}
c_{3} & = \langle \delta m_{i}^{y} \delta m_{j}^{y} \rangle_{0}~\qquad (i, j~
\text{nearest-neighbours})\\
& = \frac{1}{12} \sum_{\lambda = \pm 1} \int \frac{{\rm d}^{2}k}{\Omega} \frac{\lt(3  - \lt(1 -
\frac{2}{3} b\rt) \lambda |f_{\bk}|\rt)\lambda |f_{\bk}|}{\epsilon_{\lambda}(\bk)}~.
\end{split}
\end{equation}

The last diagram~\ref{diagrams}c, differently from $\Sigma_{\rm a}$ and $\Sigma_{\rm b}$
produces a frequency-dependent self energy term.
We find (see Appendix~\ref{ap_sigma_c}):
\begin{equation} \label{sigma_c}
\begin{split}
& \Sigma_{{\rm c} \lambda \lambda'}^{\alpha \beta}(\bk, \omega) =
\text{\includegraphics[scale=1, trim=0 1cm 0 0]{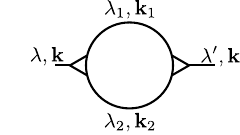}}\\
& ~ \\
& ~ \\
& =-\frac{1}{8}  \sum_{\lambda_{1}, \lambda_{2}} \int_{{\rm BZ}}  \frac{{\rm d}^{2}k_{1}}{\Omega}
\Bigg\{ \\
& \Bigg[ \frac{\Lambda^{\alpha \lambda \mu}_{\lambda \lambda_{1} \lambda_{2}}(\bk, -\bk_{1},
-\bk_{2})\Lambda_{\lambda_{1} \lambda_{2} \lambda'}^{\nu \rho \beta}(\bk_{1}, \bk_{2},
-\bk)}{\epsilon_{\lambda_{1}}(\bk_{1}) \epsilon_{\lambda_{2}}(\bk_{2})} \\
& \times \bigg[\frac{(N_{\lambda_{1}}^{\lambda \nu}(\bk_{1}) - \epsilon_{\lambda_{1}}(\bk_{1})
\sigma_{y}^{\lambda \nu}) (N_{\lambda_{2}}^{\mu \rho}(\bk_{2}) - \epsilon_{\lambda_{2}}(\bk_{2})
\sigma_{y}^{\mu \rho})}{\omega - \epsilon_{\lambda_{1}}(\bk_{1}) -
\epsilon_{\lambda_{2}}(\bk_{2}) + i0^{+}}\\
& - \frac{(N_{\lambda_{1}}^{\lambda \nu}(\bk_{1}) + \epsilon_{\lambda_{1}}(\bk_{1})
\sigma_{y}^{\lambda \nu}) (N_{\lambda_{2}}^{\mu \rho}(\bk_{2}) + \epsilon_{\lambda_{2}}(\bk_{2})
\sigma_{y}^{\mu \rho})}{\omega + \epsilon_{\lambda_{1}}(\bk_{1}) + \epsilon_{\lambda_{2}}(\bk_{2})
- i0^{+}}\Bigg] \Bigg] \Bigg\}~,
\end{split}
\end{equation}
where $\bk_{2} = \bk - \bk_{1}$, and $\Lambda$ is the vertex function
\begin{equation} \label{Lambda_vertex}
\begin{split}
& \Lambda^{\alpha \beta \gamma}_{\lambda_{1} \lambda_{2} \lambda_{3}}(\bk_{1}, \bk_{2}, \bk_{3}) =
-\frac{i}{2} \sqrt{\frac{V(1 - b/3)}{|f_{\bk_{1}} f_{\bk_{2}} f_{\bk_{3}}|}} \\
& \times \lt(\lambda_{1} \lambda_{2} \lambda_{3} \sqrt{f^{*}_{\bk_{1}}} \sqrt{
f^{*}_{\bk_{2}}} \sqrt{f^{*}_{\bk_{3}}} - \sqrt{f_{\bk_{1}}} \sqrt{f_{\bk_{2}}}
\sqrt{f_{\bk_{3}}}\rt)\\
&  \times (\lambda_{1} |f_{\bk_{1}}| \delta^{\alpha 1} \delta^{\beta \gamma} + \lambda_{2}
|f_{\bk_{2}}| \delta^{\beta 1} \delta^{\gamma \alpha} + \lambda_{3} |f_{\bk_{3}}| \delta^{\gamma
1} \delta^{\alpha \beta})~.
\end{split}
\end{equation}

In the momentum integral in Eq.~\eqref{sigma_c}, $\bk_{1}$ runs over the first Brillouin zone,
whereas, $\bk_{2}$ is everywhere given by $\bk_{2} = \bk - \bk_{1}$, irrespective of
whether $\bk_{2}$ falls within the first zone.
(As before, $\sqrt{f}$ stands everywhere for the square roots with positive real part).

The terms in the last two lines of Eq.~\eqref{sigma_c} represent, respectively, retarded
and advanced contributions to the self-energy.
We note that the factors $N(\bk) - \epsilon(\bk) \sigma_{y}$ and $N(\bk) + \epsilon(\bk) \sigma_{y}$
can be expressed in a diagonal form using Eq.~\eqref{residues}.

\section{Pseudo-Goldstone gap at zero magnetic field}
\label{pseudo-Goldstone_gap}

Using the explicit form of the self-energy we can now discuss the correction to the spectrum.
We discuss first the spectrum at $\bk = 0$ and $b = 0$.
In this case, the self-energy corrections have a strong effect: they provide a gap to the
pseudo-Goldstone mode.

For $b = 0$, the spectrum presents a gapless pseudo-Goldstone mode at long wavelengths, associated
to the degeneracy of the classical energy~\eqref{E_cl} under the rescaling $\varphi_{i} \to \xi_{i}
\varphi_{i}$, (with $\xi_{i} = \lambda$, $1/\lambda$ on the two sublattices).
The order-by-disorder mechanism however selects the Y state as a stable ground state; this state
has the lowest zero-point energy in linear spin-wave theory.
The quantum order-by-disorder selection is inevitably associated with the generation of a
gap to pseudo-Goldstone fluctuation~\cite{murthy_prb_1997, rau_prl_2018}.

To calculate the gap, we consider the Dyson equation at zero momentum and low frequency, for $b =
0$.
To make the physical interpretation more transparent, we find it convenient to return temporarily
to the sublattice basis, i.e.~to express $G_{0}^{-1}$ and $\Sigma$ as matrices with sublattice
indices $\tau$, $\tau'$ instead of branch indices $\lambda$, $\lambda'$.
At the non-interacting level, the inverse of the Green function for $\bk = 0$ reads:
\begin{equation}\label{Dyson_k=0}
\begin{split}
G^{-1 \alpha \beta}_{0\tau \tau'}(0, \omega)  & = - 3 \big(\delta_{\tau \tau'}\delta^{\alpha \beta}
- \tilde{\sigma}_{x \tau \tau'} \sigma_{z}^{\alpha \beta}\big) \\
& - (1 + i 0^{+}) \omega \delta_{\tau \tau'} \sigma_{y}^{\alpha \beta} ~,
\end{split}
\end{equation}
where $\tilde{\sigma}_{x} = \begin{pmatrix}
                             0 & 1 \\
                             1 & 0
                            \end{pmatrix}
$ is the $x$ Pauli matrix, acting in sublattice space.
Here and in the following, we will use the symbol $\tilde{\sigma}_{\mu}$ to denote Pauli matrices in
sublattice space, and $\sigma_{\mu}$ for matrices in cartesian space.

The bare spectrum is determined by the values of $\omega$ for which $\det G_{0}^{-1} = 0$.
The matrix $G_{0}^{-1}$ has two gapless modes with $\omega = 0$.
The two modes can be written, schematically, as the vectors $|\tilde{\sigma}_{x} = \sigma_{z} =
-1\rangle$, $|\tilde{\sigma}_{x} = \sigma_{z} = +1 \rangle$.
By this notation we mean the eigenvectors of $\tilde{\sigma}_{x}$ and $\sigma_{z}$ with eigenvalues
$+1$ or $-1$.

The mode $|\tilde{\sigma}_{x} = \sigma_{z} = -1\rangle$ describes transverse fluctuations with
opposite amplitudes on the two sublattices, and can be identified with the Goldstone mode,
related to the broken $U(1)$ symmetry of the model.
The mode $|\tilde{\sigma}_{x} = \sigma_{z} = +1\rangle $ instead describes longitudinal
fluctuations, with equal amplitudes on the two sublattices.
This corresponds to the pseudo-Goldstone direction, which is gapless only at zero order, due to the
accidental degeneracy of the classical limit.

To determine the spectrum, we consider the solution of the Dyson equation equation at zero momentum
\begin{equation}
G^{-1}(0, \omega) =  -3(1 - \tilde{\sigma}_{x} \sigma_{z}) - \omega \sigma_{y} -
\Sigma(0, \omega)
\end{equation}
treating formally the self-energy $\Sigma$ as a perturbation.
The analysis which we use is analogue to that presented in Ref.~\cite{rau_prl_2018}.
However, we employ a different basis, and we present additional considerations on the properties
of the self-energy.

For small $V$, we can assume that the energy of the pseudo-Goldstone gap remains small, and thus
that the self-energy $\Sigma(0, \omega)$ in Eq.~\eqref{Dyson_k=0} can be replaced with $\Sigma(0,
0)$.
Using the results of Sec.~\ref{self_energy}, we find that the self-energy at zero frequency,
momentum, and field is given formally by:
\begin{equation}\label{sigma_00}
\begin{split}
\frac{1}{V} \Sigma(0, 0) & = \frac{1}{4} (9 I_{-1} + I_{1} - 6) \tilde{\sigma}_{x} (1 +
\sigma_{z})\\
& - \frac{1}{2} \lt(81 I_{-3} - 18  I_{-1} + I_{1} \rt) (1 - \tilde{\sigma}_{x}
\sigma_{z}) \\
& + \frac{1}{4} (9 I_{-1} - I_{1}) (1 + \sigma_{z})~,
\end{split}
\end{equation}
where $I_{1} = \langle \epsilon(\bk)\rangle_{\bk}$, $I_{-1} = \langle
1/(\epsilon(\bk))\rangle_{\bk}$, $I_{-3} = \lt \langle 1/(\epsilon^{3}(\bk))
\rt\rangle_{\bk}$ are averages over the Brillouin zone of $\epsilon$, $\epsilon^{-1}$, and
$\epsilon^{-3}$, respectively.
Here $\epsilon(\bk) = \sqrt{9 - |f_{\bk}|^{2}}$ is the bare dispersion relation at zero field.

To determine the dispersion, we can treat both $\Sigma$ and $\omega$ as perturbations in the
matrix Dyson equation.
From a simple analysis we see that $\omega^{2}$ is of the same order as $\Sigma$ for $\bk = 0$, due
to the fact that the bare energy vanishes.
Thus terms of order $\omega^{2}$ have to be included in the analysis.

With this consideration we find after an explicit calculation that the spectrum is given by:
\begin{equation}
\frac{1}{6}\omega_{\lambda}^{2} - v_{0 \lambda}^{+} \Sigma(0, 0) v_{0 \lambda} = 0~,
\end{equation}
where $v_{0 \lambda}$ are the zero-order solutions,~i.e. the solutions of $G_{0}^{-1}(0,
0) v_{0 \lambda} = 0$.

For the Goldstone mode, the solution $v_{0 -1}$ is the vector
$|\tilde{\sigma}_{x} = \sigma_{z} = -1\rangle$.
It is immediate to see that all terms in Eq.~\eqref{sigma_00} vanish for
$\tilde{\sigma}_{x} = \sigma_{z} = -1$.
Thus, $v_{0-1}^{+} \Sigma(0, 0) v_{0 1} = 0$ and the Goldstone mode remains gapless.

For the pseudo-Goldstone mode instead $v_{01}$ is the vector $|\tilde{\sigma}_{x} = \sigma_{z} =
1\rangle$.
Then we find that the pseudo-Goldstone mode acquires a gap
\begin{equation} \label{pseudo-Goldstone}
\epsilon_{\rm g} = \sqrt{18 V (3 I_{-1} - 1)} \simeq 3.05 \sqrt{V}~.
\end{equation}

The gap is a nonanalytic function of $V$ and thus of $S$.
Reintroducing a factor $S J_{xy}$ and expressing the result in standard units of
measurement, we find a gap of order $3.05 \sqrt{S \alpha} J_{zz}$.
The gap thus scales as $S^{1/2}$, consistently with earlier analyses
~\cite{murthy_prb_1997, rau_prl_2018}.

Before continuing the discussion, an important remark is in order.
The self-energy $\Sigma(0, 0)$, which was used in the derivations, is actually only defined
formally because the momentum integral $I_{-3}$ is infrared (IR) divergent.
The divergence however is proportional to the matrix $(1 - \tilde{\sigma}_{x}\sigma_{z})$ which
vanishes when projected along the vectors $v_{0 \lambda}$.
This is why the result for the pseudo-Goldstone gap~\eqref{pseudo-Goldstone} is well defined, and
depends only on IR convergent integrals.

The IR divergence derives from the $\bk_{1} \to 0$ region in the integral for the self-energy
$\Sigma_{{\rm c}}(0, 0)$.
It can be checked that both Goldstone and pseudo-Goldstone modes contribute to the divergence.
For example, it is simple to verify that $\Sigma(0, 0)$ remains divergent at $b > 0$, when only
Goldstone fluctuations are divergent.

\section{Self-consistent theory of the spectrum at zero field}
\label{corrections_to_the_spectrum}

We now turn our attention to the spectrum at $\bk \neq 0$.
The analysis of this case encounters difficulties because infrared divergences induce a breakdown
of the on-shell approximation.
In fact, within the first-order on-shell approximation  the dispersion relation and the damping
rates are determined respectively by the real and the imaginary parts of the complex spectrum
\begin{equation} \label{on_shell}
\begin{split}
\omega_{\lambda}(\bk)  & = \epsilon_{\lambda}(\bk) \\
& + \frac{3 + \frac{b}{3} \lambda
|f_{\bk}|}{\epsilon_{\lambda}(\bk)}  v_{0\lambda}^{+}(\bk) \Sigma(\bk, \epsilon_{\lambda}(\bk))
v_{0\lambda}(\bk)~,
\end{split}
\end{equation}
where
\begin{equation} \label{v}
v_{0\lambda}(\bk) = \frac{1}{\sqrt{2 \lt(3 + \frac{b}{3} \lambda |f_{\bk}| \rt)}}
w^{\alpha}_{\lambda} (\bk)~,
\end{equation}
and $w^{\alpha}_{\lambda}(\bk)$ are the vectors introduced in Eq.~\eqref{w_vectors}.
The diagonal element $v_{0\lambda}^{+}(\bk) \Sigma(\bk, \epsilon_{\lambda}(\bk)) v_{0\lambda}(\bk)$
corresponds to the ``normal'' part of the self-energy for the branch $\lambda$.
(The anomalous parts, instead, do not contribute to the corrections of the spectrum at leading
order~\cite{chubukov_jpcm_1994}).

As $b \to 0$, Eq.~\eqref{on_shell} diverges at all values of the momentum $\bk$ (apart from the
$\Gamma$ and the $K$ points).
The divergences arise from the regions $|\bk_{1}| \ll 1$ and $|\bk_{2}| \ll 1$ in the self-energy
diagram $\Sigma_{{\rm c}}$, and receive contributions from both Goldstone and pseudo-Goldstone
modes.
The soft Goldstone modes with $|\bk_{1}| \ll 1$ and $|\bk_{2}| \ll 1$ produce a divergence of the
matrix self-energy $\Sigma(\bk, \omega)$ at $\omega =\epsilon_{\lambda}(\bk)$.
This divergence however cancels in the diagonal matrix element $\Sigma_{\lambda}$
(see App.~\ref{IR_divergence}).
Thus, the soft Goldstone modes are not problematic in the first-order calculation of the
energy spectrum.
We attribute this IR divergence to the fact that the single-magnon dispersion coincides with
branch-cut singularity of the spectral function.
The singularity can then be traced to a difficulty in defining the residue of the single-magnon
pole.

On the other hand, the soft pseudo-Goldstone modes generate a IR divergence for $b = 0$ which does
not cancel in the matrix element $\Sigma_{\lambda}$.
The divergence can be studied expanding the expression for $\Sigma_{\rm c}$ in the $\bk_{1} \to 0$
region.
From the explicit expression of $\Sigma_{\rm c}$, Eq.~\eqref{sigma_c}, we find that the
soft pseudo-Goldstone fluctuations give a contribution to the self-energy which is independent of
the branch index $\lambda = \pm 1$ and which has the form:
\begin{equation} \label{Sigma_div}
\begin{split}
& \Sigma_{\lambda}^{(\bk_{1} \to 0)}(\bk, \omega) = 9 V \epsilon(\bk) \int_{|\bk_{1}| \ll
1}\frac{{\rm d}^{2}k_{1}}{\Omega} \\
& \qquad \times \frac{1}{\epsilon(\bk_{1}) (\omega - \epsilon(\bk_{1}) - \epsilon(\bk - \bk_{1}) +
i0^{+})}~.
\end{split}
\end{equation}

This term has a singularity on the ``mass shell'' $\omega = \epsilon(\bk)$.
In particular, the imaginary part presents a finite jump and the real part a logarithmic divergence
at $\omega = \epsilon(\bk)$.
This makes it impossible to define the on-shell approximation at any momentum at $b = 0$.

This IR divergence arises from the gapless spectrum of the pseudo-Goldstone mode.
In the case of ``true'' Goldstone mode, the gapless spectrum does not induce a divergent correction
because the vertex function vanishes at low momentum, compensating for the singular propagator.
The vanishing of the vertex function is a a consequence of the broken $U(1)$ symmetry and of the
related Ward identities.
For pseudo-Goldstone modes, instead, the gapless spectrum is not compensated by a vanishing matrix
element, as there is no exact symmetry and Ward identity associated with it.
As a result, a divergence is generated in the one-loop correction.

The resolution of this problem rests on the fact that the physical pseudo-Goldstone mode is, in
fact, gapped by the order-by-disorder mechanism~\cite{rau_prl_2018, murthy_prb_1997}, as discussed
in Sec.~\ref{pseudo-Goldstone_gap}.
The gap directly removes the problem in Eq.~\eqref{Sigma_div}, regularizing the denominator
$1/\epsilon(\bk_{1})$ in Eq.~\eqref{Sigma_div}.
Thus the imaginary part of $\Sigma$ does not present a finite jump.
In addition, the emission of a virtual pseudo-Goldstone mode with momentum $\bk_{1}$ costs an energy
$\epsilon_{\rm g}$, and thus the energy denominator $\omega - \epsilon(\bk_{1}) - \epsilon(\bk -
\bk_{1}) + i 0^{+}$ does not vanish.
Instead of a divergence, the region $\bk_{1} \to 0$ produces a purely real (non-resonant)
contribution, which depends logarithmically on $\epsilon_{g}$.
For $V$ small, since $\epsilon_{\rm g} \simeq 3.05 \sqrt{V}$, we obtain a non-analytic correction of
order $V \ln 1/\sqrt{\epsilon_{\rm g}} \propto - V \ln V$ to the self-energy.

Thus, it is essential to consider a renormalized propagator, which encodes the presence of a
pseudo-Goldstone gap in the internal lines of the diagrams.
In this work we use an approximation based on the renormalization of two parameters.
We consider in particular the introduction of a renormalized magnetic field $\tilde{b}$, and an
energy-scale renormalization $z$.
In other words, we assume that the calculation of the spectrum at zero physical field can be
approximated using the one-loop self-energy calculated at a finite $\tilde{b}$, after an energy
rescaling.
This approximation is automatically consistent with the Ward identities and with the gaplessness
of the Goldstone mode, because the field term $\tilde{b} \sum_{i} a^{+}_{i} a_{i}$ is
$U(1)$-symmetric and is naturally generated by renormalization.
The same is true with the energy-scale renormalization, which is equivalent to a time dilation,
and is therefore consistent with symmetries.
The renormalization of the field $\tilde{b}$ provides a gap to the pseudo-Goldstone mode, which is
crucial to obtain finite results.
Here, we introduce a field $\tilde{b}$ tuned self-consistently in order to match the gap
$\epsilon_{\rm g}$ at $\bk = 0$.
The energy renormalization $z$, instead, is not strictly necessary to remove the IR divergences.
However, we introduced it because, as shown below, one of the significant effects of the
interactions is a reduction of the bandwidth.
As it will be shown in the following pages, the simple renormalization of $\tilde{b}$ and $z$ does
not fully capture some features of the spectrum, in particular a tendency of the corrections to
induce a crossing of the two branches.
However, in this work we confine to this approximation for simplicity.

Within this approximation the renormalized equations and the spectrum have a simple form.
The Dyson equation at zero physical field becomes replaced by
\begin{equation}
G^{-1}(\bk, \omega; 0) = G_{0}^{-1}(\bk, \omega; 0) - z\Sigma(\bk, z^{-1} \omega; \tilde{b})~,
\end{equation}
where $G^{-1}_{0}(\bk, \omega; 0)$ is the bare propagator at zero field and $\Sigma(\bk, \omega; b)$
is the one-loop self-energy at finite magnetic field $b$, calculated in the previous sections.
(Here and in the following we express explicitly the magnetic field dependence of all quantities
involved in the calculations).

The expression $z\Sigma(\bk, z^{-1} \omega; \tilde{b})$ coincides with the self-energy computed in
a renormalized model in which the magnetic field is equal to $\tilde{b}$, and the energy scale is
renormalized by a factor $z$.
The spectrum of the renormalized model is $\omega = z \epsilon_{\lambda}(\bk, \tilde{b})$.
If $G_{0}^{-1}(\bk, \omega; 0) - z\Sigma(\bk, z^{-1} \omega; \tilde{b})$ was equal to $z
G_{0}^{-1}(\bk, \omega/z; \tilde{b})$, the approximation would be exactly self-consistent and the
spectrum would be exactly equal to $\omega = z\epsilon_{\lambda}(\bk, \tilde{b})$ at all momenta
$\bk$.

The simplest way to compute the spectrum is an analogue of the first-order on-shell approximation.
Taking into account the renormalizations, this approximation can be derived by treating formally $z
G_{0}^{-1}(\bk, \omega/z; \tilde{b})$ as the zero-order term and $G^{-1}(\bk, \omega, 0) -
zG_{0}^{-1}(\bk, \omega/z; \tilde{b})$ as a first-order perturbation.
Assuming that $\omega \simeq z \epsilon_{\lambda}(\bk, \tilde{b})$ up to small corrections, we can
replace $\omega = z \epsilon_{\lambda}(\bk, \tilde{b})$ in the argument of $z \Sigma(\bk, \omega/z;
\tilde{b})$.
As shown in App.~\ref{renormalized_perturbation_theory}, we obtain for the two branches at zero
physical field:
\begin{equation}\label{spectrum}
\begin{split}
\omega_{\lambda}(\bk; 0) & = \frac{9 - |f_{\bk}|^{2} + \frac{\tilde{b}}{3} \lt(3 + \lambda
|f_{\bk}|\rt)\lambda |f_{\bk}|}{\epsilon_{\lambda}(\bk)} \\
& +  \frac{z \lt(3 + \frac{\tilde{b}}{3} \lambda |f_{\bk}|\rt)
\Sigma_{\lambda}(\bk, \tilde{b})}{\epsilon_{\lambda}(\bk, \tilde{b})}~,
\end{split}
\end{equation}
where $\Sigma_{\lambda}(\bk, \tilde{b}) = v_{0 \lambda}^{+}(\bk) \Sigma(\bk,
\epsilon_{\lambda}(\bk, \tilde{b}))
v_{0 \lambda}(\bk)$ is the on-shell self-energy calculated at a field $\tilde{b}$.

The renormalization parameters $\tilde{b}$ and $z$ can be fixed via self-consistent relations in
such way that the spectrum is as close as possible to being self-consistent,~\emph{i.e.} in
such way that $\omega_{\lambda}(\bk, 0)$ is close to $z \epsilon_{\lambda}(\bk, \tilde{b})$.
Here, we fix $\tilde{b}$ and $z$ imposing two self-consistency relations.
First, we require the pseudo-Goldstone gap to be exactly self-consistent at the $\Gamma$ point.
This leads to the condition $\omega_{+1}(0) = z \sqrt{12 b}$, which is equivalent to
\begin{equation} \label{ren_1}
12 z b = 6 b + z (3 + b) {\rm Re} \Sigma_{+1}(0, b)~.
\end{equation}

As a second condition, we impose the matching of the dispersion at the $Y$ point.
Since $f_{\bk} = 0$ and $\epsilon_{\lambda}(\bk) = 3$ at this point, the matching condition
requires:
\begin{equation} \label{ren_2}
3z = 3 + z {\rm Re} \Sigma_{\lambda}(Y, \tilde{b})~.
\end{equation}

This second condition is consistent because at the $Y$ point $\Sigma_{\lambda}$ does not depend on
the branch index.

The approximation expressed in Eq.~\eqref{spectrum} neglects a characteristic feature of the bare
spectrum at $b= 0$: the fact that for zero magnetic field the dispersion relations of the two
branches $\lambda = \pm 1$ are exactly degenerate at all momenta $\bk$ in the Brillouin zone.
In other words, Eq.~\eqref{spectrum} assumes that the interactions split the degeneracy in the same
way as an external field does.
This is certainly true at $\bk = 0$, where the branches have the character of Goldstone and
pseudo-Goldstone fluctuations.
However, for generic $\bk$, the interactions may lift the degeneracy differently, selecting
superpositions of the $|\lambda=\pm 1, \bk\rangle$ states as the relevant magnon eigenstates.

To assess this effect, we consider a matrix generalization of Eq.~\eqref{spectrum}, which includes
off-diagonal matrix elements of the self-energy between different magnon branches.
If we ignored the IR divergences and computed the spectrum directly in the on-shell approximation
for $b = 0$, the dispersions would be given formally by the frequencies $\omega$ at which the $2
\times 2$ matrix:
\begin{equation}\label{off_diag_sigma}
v_{0\lambda}^{+}(\bk, 0)  (G_{0}^{-1}(\bk, \omega) - \Sigma(\bk, \epsilon(\bk, 0)))
v_{0\lambda'}(\bk, 0)
\end{equation}
has vanishing eigenvalues.
The matrix $\Sigma_{\lambda \lambda'}(\bk, 0) =  v_{0\lambda}^{+}(\bk, 0)\Sigma(\bk, \epsilon(\bk,
0))v_{0\lambda'}(\bk, 0)$ is a matrix of ``normal'' self-energies, including hybridization between
different modes.

To include the renormalizations, which are essential to regularize the IR divergences, we
consider
the zero eigenvalues of the $2 \times 2$ matrix equation
\begin{equation}\label{off_diag_sigma_2}
v_{0\lambda}^{+}(\bk, \tilde{b})  (G_{0}^{-1}(\bk, \omega) - z\Sigma(\bk, z^{-1} \omega, \tilde{b}))
v_{0\lambda'}(\bk, \tilde{b})~.
\end{equation}

In contrast with the diagonal matrix elements, the off-diagonal elements $\Sigma_{\lambda
\lambda'}(\bk; 0) = v_{0 \lambda}^{+}(\bk) \Sigma (\bk, \epsilon(\bk, 0); 0) v_{0 \lambda'}(\bk)$
with $\lambda \neq \lambda'$ do not present infrared divergences on-shell, and
could be calculated consistently at $b = 0$.
In this work, however, we calculated the off-diagonal elements $\Sigma_{+1, -1}$ and $\Sigma_{-1,
+1}$ at the same value of the renormalized field $\tilde{b}$ used in the calculation of the diagonal
elements.

In the definition of the on-shell approximation, an ambiguity arises as to which frequency $\omega$
to consider as the ``on-shell'' one.
Because the two dispersions $\epsilon_{\lambda}(\bk, \tilde{b})$ have different energies at
$\tilde{b} \neq 0$ the ``on-shell'' frequency corresponding to an off-diagonal matrix elements is
not strictly defined.
In this work, we choose to replace, for the off-diagonal self-energy elements, the bare spectrum
calculated at zero field.
This resolves the ambiguity because $\epsilon_{\lambda}(\bk, 0)$ is independent of the branch index.

The magnon branches can then be calculated from the eigenvalues of the $2 \times 2$ matrix (see
App.~\ref{renormalized_perturbation_theory}):
\begin{equation} \label{C_k}
\begin{split}
& C_{\lambda \lambda'}(\bk)  =  \frac{9 - |f_{\bk}|^{2} + \frac{\tilde{b}}{3} (3 +
\lambda |f_{\bk}|) \lambda |f_{\bk}|}{\epsilon_{\lambda}(\bk, \tilde{b} )} \delta_{\lambda
\lambda'} \\
& + z \sqrt{\frac{\lt(3 + \frac{\tilde{b}}{3} \lambda |f_{\bk}|\rt)\lt(3 + \frac{\tilde{b}}{3}
\lambda' |f_{\bk}|\rt)}{\epsilon_{\lambda}(\bk, \tilde{b}) \epsilon_{\lambda'}(\bk, \tilde{b})}}
\Sigma_{\lambda \lambda'}(\bk, e_{\lambda \lambda'}(\bk, \tilde{b}); \tilde{b})~,
\end{split}
\end{equation}
where $e_{\lambda \lambda'}(\bk) = \delta_{\lambda \lambda'} \epsilon_{\lambda}(\bk, \tilde{b}) +
(1 - \delta_{\lambda \lambda'}) \epsilon(\bk, 0)$.

The off-diagonal matrix elements vanish along the $\Gamma-M$ line, as a consequence of the
reflection symmetry $x \to -x$.
Thus, the spectra computed from Eq.~\eqref{spectrum} and from Eq.~\eqref{C_k} coincide along this
line.
As a consequence, the self-consistent values of $\tilde{b}$ and $z$ are the same in the two
approximations.
As discussed in the next section, we find that the off-diagonal terms have a weak effect also along
the $\Gamma-K$ direction.

\subsection{Numerical results and discussion}
\label{numerical_results}

To calculate the corrections to the spectrum explicitly, we computed the on-shell self-energies by
numerical integration.
For the graphs $\Sigma_{{\rm a}}$ and $\Sigma_{{\rm b}}$, corresponding to the
graphs~\ref{diagrams}a,~\ref{diagrams}b we used an adaptive integration method.
For $\Sigma_{{\rm c}} $ we used a discrete summation over a mesh of $1.6 \times 10^{7}$
$\bk$ points, replacing the infinitesimals $0^{+}$ in the energy denominators with a
small positive $\delta$.
We used $\delta = 5 \times 10^{-3}$ in all calculations, a part from those at $\tilde{b} = 0.0001$
where we used $\delta = 2 \times 10^{-3}$.
We found it convenient to calculate the self-energy term $\Sigma_{\rm c}$ by using
Eq.~\eqref{residues} and by contracting the vectors $w^{\alpha}_{\lambda}$ with the vertex function
$\Lambda$.
This gives expressions equivalent to those which would be obtained using perturbation theory after
a Bogoliubov transformation.

In the integration, we used a grid of $\bk$ points defined in polar coordinates (equally spaced
with respect to radial and angular variables, as in a spider net).
In addition, we restricted the summation to the points for which $\epsilon_{\lambda_{1}}(\bk_{1}) <
\epsilon_{\lambda_{2}}(\bk_{2})$ and multiplied the result by 2, taking advantage of the symmetry of
Eq.~\eqref{sigma_c} under the interchange $1 \leftrightarrow 2$.
This choice improves the accuracy of the integration in the region $|\bk_{1}| \ll 1$ and excludes
the region $|\bk_{2}| \ll 1$ from the integration.

\begin{figure}[t]
\centering
\includegraphics[scale=1]{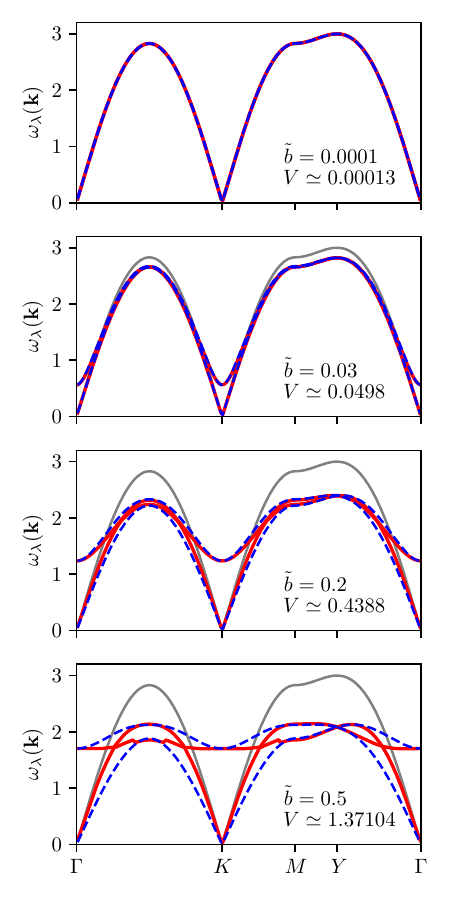}
\caption{\label{_spectrum_0_0001_0_03_0_2_0_5}
Red solid lines: dispersion relation at zero physical field within the renormalized one-loop
approximation, Eq.~\eqref{C_k}, for growing values of the coupling constant $V$, corresponding
to the renormalized fields $\tilde{b} = 0.0001, 0.03, 0.2, 0.5$.
Blue dashed lines: dispersions $z\epsilon_{\lambda}(\bk, \tilde{b})$.
Grey lines: bare spectrum at $b = 0$, $\epsilon(\bk) = \sqrt{9 - |f_{\bk}|^{2}}$.
}
\end{figure}

\begin{figure}[t]
\centering
\includegraphics[scale=1]{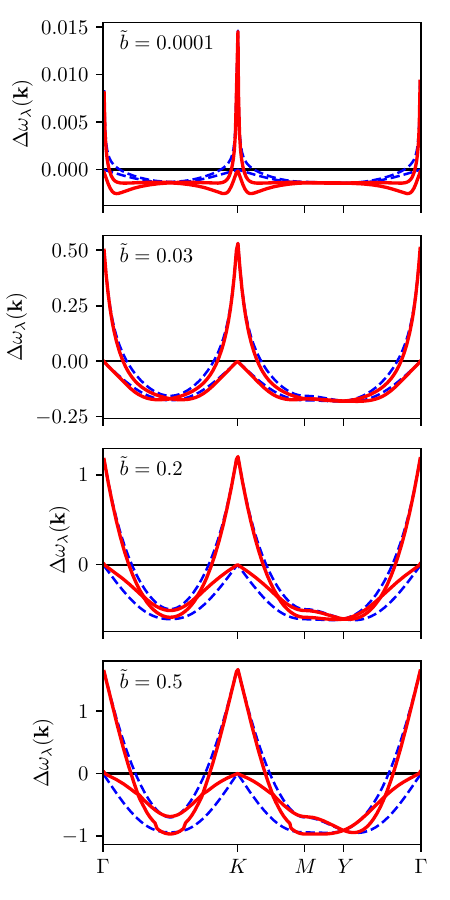}
\caption{\label{_corrections_0_0001_0_03_0_2_0_5}
Corrections to the dispersion relations at zero physical field at growing values of $V$.
Red solid lines: corrections $\omega_{\lambda}(\bk) - \epsilon(\bk, 0)$, calculated from
Eq.~\eqref{spectrum}.
Blue dashed lines: differences $z\epsilon_{\lambda}(\bk, b) - \epsilon(\bk, 0)$, with $\epsilon(\bk,
0) = \sqrt{9 - |f_{\bk}|^{2}}$ the LSWT spectrum at zero field and $z\epsilon_{\lambda}(\bk, b)$,
the LSWT spectrum in a field $b$, after bandwidth renormalization.
}
\end{figure}

\begin{figure}[t]
\centering
\includegraphics[scale=1]{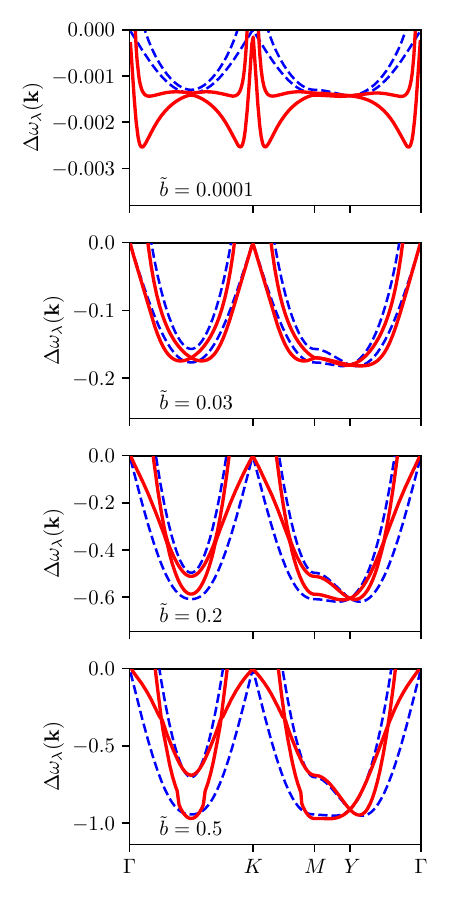}
\caption{\label{_corrections_0_0001_0_03_0_2_0_5_b}
Same plots reported in Fig.~\ref{_corrections_0_0001_0_03_0_2_0_5}, but presented showing only the
$\Delta \omega < 0$ region.
}
\end{figure}

\begin{figure}
\centering
\includegraphics[scale=1]{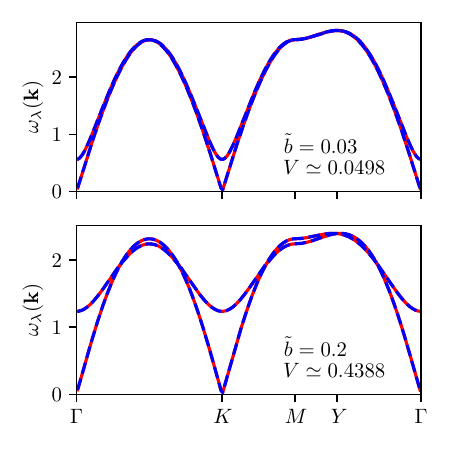}
\caption{\label{_with_and_without_mix_0_03_0_2} Red solid lines: spectra calculated with
Eq.~\eqref{C_k} (including mixing of the modes $\lambda = \pm 1$).
Blue dashed lines: spectra calculated with Eq.~\eqref{spectrum} (neglecting the mixing of the modes
$\lambda = \pm 1$).
The top and bottom panels show, respectively, the results for $\tilde{b} = 0.03$ and $\tilde{b} =
0.2$.
The difference between the spectra is not noticeable on the scale of the figure showing that the
off-diagonal mixing has a small effect for most momenta in the high-symmetry lines.}
\end{figure}

The results of the calculations are illustrated in
Figs.~\ref{_spectrum_0_0001_0_03_0_2_0_5}---~\ref{_with_and_without_mix_0_03_0_2}.
In Fig.~\ref{_spectrum_0_0001_0_03_0_2_0_5}, we show the dispersions $\omega_{\lambda}(\bk)$,
calculated from Eq.~\eqref{C_k}, together with the bare spectrum $\epsilon(\bk, 0)$ and the
dispersions $z \epsilon_{\lambda}(\bk, \tilde{b})$, for growing values of the renormalized field:
$\tilde{b} = 0.0001, 0.03, 0.2$, and $0.5$ (the physical field is zero in all cases considered).
These correspond to the coupling constants and bandwith renormalizations reported in
table~\ref{tab}.
\begin{table}
\caption{\label{tab} Values of $V$, $\tilde{b}$, and $z$, used in numerical calculations of the
spectrum. For each value of $\tilde{b}$, the corresponding values of $V$ and $z$ are deduced via
Eqs.~\eqref{z_A}---\eqref{V}.}
\begin{tabular}{ccccc}
\hline
$\qquad V\qquad $ & $1.3 \times 10^{-4}$ & $0.0498 $ & $0.4388$ & $1.371$ \\
\hline
$\qquad\tilde{b}\qquad$ &$\quad 10^{-4}\quad$ &$\quad 0.03 \quad$  &
$\qquad 0.2 \qquad$ & $\quad 0.5 \quad$  \\
$ z $ & $0.9995$ & $0.9399$ & $0.7981$ & $ 0.6954$\\
\hline
\end{tabular}
\end{table}

For sufficiently small coupling (the top three panels of Fig.~\ref{_spectrum_0_0001_0_03_0_2_0_5}),
we see that the corrected spectrum is close to $z \epsilon_{\lambda}(\bk, \tilde{b}) -
\epsilon(\bk, 0)$, in the high-symmetry directions which we studied.
This corroborates, in a first approximation, the self-consistent approach which we used.
At the same time, we note that the corrections present a nontrivial evolution as a function of $V$,
and features which are not captured by the dispersions $z \epsilon_{\lambda}(\bk, \tilde{b})$.

As $V$ grows to larger values, the deviations from self-consistency become more pronounced.
For $V \simeq 1.4$, the numerical results show marked deviations from the dispersions $z
\epsilon_{\lambda}(\bk, \tilde{b})$.
The large corrections however indicate that the first order approximation becomes invalid at $V
\simeq 1$.

Let us discuss in more details some properties of the spectrum.
In the region $V \to 0$, the corrections include different effects which scale differently with $V$.
The strongest effect is the generation of the pseudo-Goldstone gap, which scales as $\sqrt{V}$.
The spectrum at $\bk \neq 0$, instead, has a weaker dependence, linear up to logarithmic
corrections.
To present in more detail the evolution of the corrections, in
Figs.~\ref{_corrections_0_0001_0_03_0_2_0_5},~\ref{_corrections_0_0001_0_03_0_2_0_5_b} we show the
differences $\omega_{\lambda}(\bk) - \epsilon(\bk, 0)$ and $z \epsilon_{\lambda}(\bk) -
\epsilon(\bk, 0)$ for the same values of the coupling parameters used in
Fig.~\ref{_spectrum_0_0001_0_03_0_2_0_5}.

In the limit of extremely small $V$ and for momenta different from the gapless points we expect
that the corrections are dominated by the logarithmic term, of order $V \ln (1/V)$.
These can be calculated analytically by expanding the integral~\eqref{Sigma_div} for $|\bk_{1}| \ll
1$.
Using the on-shell expression~\eqref{on_shell} at $b = 0$, and cutting-off the integral at
$k_{1{\rm min}} \propto \epsilon_{\rm g}$  we find, keeping the logarithmic term:
\begin{equation} \label{spectrum_div}
\begin{split}
\omega_{\lambda}(\bk) & \simeq \epsilon(\bk, 0) -  \frac{27 V}{\Omega v_{s}} \ln
\frac{\Lambda}{\epsilon_{\rm g}} \int_{0}^{2 \pi} \frac{{\rm d}\theta}{v_{s} - v(\bk) \cos \theta
}\\
& = \epsilon(\bk) - \frac{54 \pi V}{\Omega v_{s} \sqrt{v_{s}^{2} - (v(\bk))^{2}}} \ln
\frac{\Lambda}{\epsilon_{\rm g}}~.
\end{split}
\end{equation}

Here $v_{s} = 3 \sqrt{2}/2$ is the speed of magnons at the $\Gamma$ point, $v(\bk)$ is the absolute
value of the group velocity at momentum $\bk$, and $\Lambda$ is an energy cutoff of the order of 1,
below which the integrand in $\Sigma_{{\rm c}}$ can be approximated using
Eq.~\eqref{Sigma_div}.
The numerical results shown in the top panel of Fig.~\ref{_corrections_0_0001_0_03_0_2_0_5_b} show
that, even at $\tilde{b}=0.0001$, the corrections are not yet dominated by the
form~\eqref{spectrum_div}.
This can be seen in particular in the fact that the corrections are significantly different in the
two branches $\lambda = \pm 1$.
Eq.~\eqref{spectrum_div} predicts instead the same form of the leading corrections for the two
modes $\lambda = \pm 1$.
We expect that the corrections should slowly approach Eq.~\eqref{spectrum_div} for $\tilde{b} \to
0$.

The spectra presented in Fig.~\ref{_spectrum_0_0001_0_03_0_2_0_5} do not present a roton-like
minimum near the $M$ point of the Brillouin zone.
For small $V$, the absence of a minimum is trivial: the corrections are not strong enough, and the
dispersion is dominated by the bare dispersion $\epsilon(\bk, 0)$, which has a saddle point at $M$.
However, we do not observe a minimum, also at larger $V \simeq 1.4$, in the approximation scheme
used here.

Considering, instead of the spectrum, the momentum dependence of the corrections
$\omega_{\lambda}(\bk) - \epsilon(\bk, 0)$, we note that Eq.~\eqref{spectrum_div} predicts a maximum
near $M$.
The numerical results at $\tilde{b} = 0.0001$, on the other hand, show that the correction to the
Goldstone spectrum has a maximum near $M$, whereas the correction to the pseudo-Goldstone spectrum
has a saddle.
The saddle is shallow and hardly visible in the top panel of
Fig.~\ref{_corrections_0_0001_0_03_0_2_0_5_b}.
For smaller coupling we expect that the corrections to both branches presents a maximum.

From the results in
Figs.~\ref{_spectrum_0_0001_0_03_0_2_0_5}---\ref{_corrections_0_0001_0_03_0_2_0_5_b} we also note a
crucial change in the ordering of the branches.
For $\tilde{b} = 0.0001$ the two magnon bands do not present crossing points: the branch $\lambda =
+1$ has a higher energy than the branch $\lambda = -1$ for all momenta which we studied.
Instead for higher values of the couplings, corresponding to  $\tilde{b} = 0.03$,
$\tilde{b} = 0.2$, $\tilde{b} = 0.5$, we find a crossing of the two bands.

Along the $\Gamma-K$ line and the equivalent $K-M$ line the off-diagonal self-energy elements lift
the degeneracy leading to an avoided crossing.
However we find that the anticrossing of the branches is not appreciable.
This can be traced to the fact that the off diagonal self-energy elements are numerically small.
This makes the regions of avoided crossing very narrow.
Along the $\Gamma -M$ line, instead, the off-diagonal self-energy elements are strictly zero.
Thus the crossing is not avoided along this line.

The crossing of branches is not encoded in a self-consistent way in the renormalized spectrum which
we used in our calculation: the spectrum $z \epsilon_{\lambda}(\bk, \tilde{b})$ which we used as an
ansatz in the internal lines of the diagrams does not present the branch inversion which emerges
as an ``output'' of the calculation.
This shows that the approximation which we used cannot be considered as fully self-consistent at
all momenta.
The approximation which was employed is rather a ``one-shot'' first-order calculation, improved
in order to cutoff the IR divergences and to encode the bandwidth rescaling.

Before concluding the discussion, some remarks are in order.
In addition to the spectrum renormalization, the interactions lead to magnon decays, leading to a
finite magnon lifetime~\cite{chernyshev_prl_2006, chernyshev_prb_2009, maksimov_prb_2016}.
The linear spin-wave dispersions analyzed in this work are identical to the LSWT dispersions of the
honeycomb XXZ model, for which the magnon decays have been analyzed in
Ref.~\cite{maksimov_prb_2016}.
For the values of the field $\tilde{b} = 10^{-4}$---$0.5$ considered in this section, the 
only decay process which is kinematically allowed is the disintegration of one high-energy 
magnon in the branch into two softer magnons, both belonging to the lower branch.
For small $\tilde{b}$ the phase space for this decay process is small.
In fact, strictly at $\tilde{b} = 0$ the dispersion of the single-magnon excitations coincides
everywhere with the bottom of the two-magnon continuum. Thus, no decay is possible at $\tilde{b} =
0$.
For $\tilde{b} \ll 1$ the decay of the high-energy branch into two magnons becomes allowed, but has
however a small decay rate.
In our calculations we found that the imaginary part ${\rm Im}\omega$ is much smaller than the real
part ${\rm Re}\omega_{\lambda}$ at generic points in the Brillouin zone.

As a second remark, we note that in the on-shell approximation which we considered, the spectrum
must present anomalies due to van Hove singularities in the two-magnon continuum.
Following the analysis of Ref.~\cite{maksimov_prb_2016}, we deduce that the anomalies occur at
momenta $\bk$ for which
\begin{equation}\label{van_hove}
\epsilon_{\lambda = +1}(\bk, \tilde{b}) = 2 \epsilon_{\lambda = -1}(\bk/2, \tilde{b})~,
\end{equation}
\emph{i.e.} at the momenta for which a magnon can disintegrate into two magnons with the same final
momentum $\bk_{1} = \bk_{2} = \bk/2$.
In addition, since the off-diagonal matrix elements were calculated assuming $\omega =
\epsilon(\bk, 0)$ as the on-shell frequency, these elements present singularities for
\begin{equation}\label{van_hove_2}
\epsilon(\bk, 0) = 2 \epsilon_{\lambda = -1}(\bk/2, \tilde{b})~.
\end{equation}

Eq.~\eqref{van_hove},~\eqref{van_hove_2} have solution for arbitrarily small $\tilde{b}$, and
defines surfaces in momentum space in which the self-energy in the on-shell
approximation~\eqref{C_k} is singular.

By analyzing the imaginary parts of the diagonal self-energy elements we verified that these
present peaks at momenta corresponding to the solutions of~\eqref{van_hove} as expected.
In the real parts, the van Hove singularity are expected to give rise to jump-like
anomalies~\cite{chernyshev_prl_2006, chernyshev_prb_2009}.
In our numerical results singularities are clearly visible in the data calculated at $\tilde{b} =
0.5$, (see the bottom panel of Fig.~\ref{_spectrum_0_0001_0_03_0_2_0_5}).
The jumps in the spectrum seen along the $\Gamma$-$K$ and the $K$-$M$ directions are different
images of the same singularity, equivalent under reciprocal lattice translations.
Along the $\Gamma$-$M$ line, instead, the same van-Hove singularity is not visible, because it is
suppressed by the vanishing of the vertex function~\eqref{Lambda_vertex}.
In the data sets calculated at smaller $\tilde{b}$, instead, the anomalies are weaker and cannot be
clearly distinguished in the plots.
In this respect, we note that the introduction of a nonzero $\delta$ acts as a regularization of the
singularity~\cite{chernyshev_prl_2006, chernyshev_prb_2009}, leading to a smoothened spectrum.
A more detailed analysis of the singularities requires a computation beyond the one-loop
approximation, in analogy with the theory presented in Refs.~\cite{chernyshev_prl_2006,
chernyshev_prb_2009}.

Next, we comment on the role of mixing between the two branches, induced by the off-diagonal
matrix elements $\Sigma_{\lambda \lambda'}$ ($\lambda \neq \lambda'$).
Numerically, we found that the effect of the off-diagonal elements is weak, due to small values of
the off-diagonal elements of $\Sigma$.
In Fig.~\eqref{_with_and_without_mix_0_03_0_2}, we show a comparison between the spectra calculated
according to Eq.~\eqref{C_k}, and the spectra computed with the simpler Eq.~\eqref{spectrum}, which
neglects the off-diagonal mixing elements. The spectra, computed for $\tilde{b} = 0.03$ and
$\tilde{b} = 0.2$ are seen to be essentially undistinguishable, indicating that the off-diagonal
mixing elements have a minor effect (a part from leading to very narrow regions of
avoided crossing near the degeneracy points).

\subsection{Comparison with the zero-field spectrum of K$_{2}$Co(SeO$_{3}$)$_{2}$}
\label{comparison_KCSO}

In the light of the results obtained in the previous section, we can comment on the connection
with the zero-field spectrum observed experimentally in K$_{2}$Co(SeO$_{3}$)$_{2}$.
Due to the low spin $S = 1/2$ and the extremely strong anisotropy $J_{xy}/J_{zz} \simeq
0.07$, the coupling constant $V$ relevant to KCSO is very
large.
Assuming $\alpha = 1/14$~\cite{chen_nc_2026}, the coupling $V$ is as large as $V = 28$.
Thus the system lies deep in the strong coupling regime, and is much closer to the
``quantum Ising'' regime $V \to \infty$ than it is to the semiclassical regime $V \ll 1$.
For this reason, we cannot expect that the one-loop approximation gives a quantitative
theory directly applicable to study the experimental spectrum.

The results of our analysis show that, even from a qualitative point of view, the nonlinear
spin-wave expansion does not provide a simple framework in which to interpret the spectral features
observed experimentally.
In particular, the experimental spectrum measured at zero field in KCSO reveals a pronounced
roton-like minimum at the $M$ points, and a low-energy gapped mode at the $K$ point, which was
suggested to reflect a pseudo-Goldstone excitation~\cite{zhu_npj_2025}.

Within the spin-wave expansion, the single-magnon spectrum can only develop a roton-like minimum if
the coupling $V$ is sufficiently strong to compensate for the curvature of the bare dispersion.
Our results, however, indicate that the inversion of the curvature at the $M$ point does not
occur, even if $V$ is pushed to moderately large values (of the order of $V \simeq 1$).

Besides, the pseudo-Goldstone gap grows rapidly as a function of $V$ (asymptotically,
as $\epsilon_{\rm g} \approx 3.05 \sqrt{V}$ for small $V$), and reaches a value comparable
with the total bandwidth of the excitation spectrum for $V \simeq 1$.

To discuss in more detail the gap of the pseudo-Goldstone excitation, we note that
Ref.~\cite{zhu_npj_2025} identified in the neutron scattering spectrum of KCSO a gapped mode with
energy $60\mu$eV, which corresponds to approximately $20$\% of the characteristic bandwidth $\Delta
E = 3 S J_{xy}$.
This mode was interpreted as the pseudo-Goldstone excitation.
Since KCSO presents an extremely strong degree of easy-axis anisotropy, this interpretation
indicates that the quantum Ising model, in the limit, $J_{zz} \to \infty$ must support a mode with
relatively low energy, with a bottom lying at approximately at an energy $0.2 \times (3 S J_{xy})$.

However, the results derived theoretically in the previous sections show that the
pseudo-Goldstone gap computed within the nonlinear spin-wave expansion grows quickly as a function
of $V$, and reaches a value comparable to $0.2 \times (3 S J_{xy})$ already at small values of the
coupling $V$.
We find in particular $\epsilon_{\rm g}/\Delta E \simeq 19$\% in the data calculated at $V \simeq
0.0498$.
Since the one-loop approximation is justified in the small $V$ region, we expect that the rapid
growth of $\epsilon_{\rm g}/(3 S J_{xy})$ as a function of $V$ is a robust prediction.
Thus, we expect that upon increasing $V$ beyond $V \approx 0.05$, the gap grows to values larger
than that observed experimentally in KCSO.

This observation suggests that the gap may behave non-monotonically as a function of $V$, in such a
way that $\epsilon_{\rm g}$ is small for $V \ll 1$ and $V \gg 1$, with larger values at
intermediate values of $V$.
Other possibilities however cannot be excluded.
One alternative scenario is that the gap observed in experiment cannot be directly explained via a
continuation of the spin-wave result.
This scenario would imply that the gapped excitation observed experimentally cannot be interpreted
directly according to the conventional picture of pseudo-Goldstone excitations, which is inherently
based on the semiclassical expansion.

A third possibility is that the fluctuation of the spins in the C sublattice, which is neglected in
the Hamiltonian~\eqref{H_ef}, plays an important role in lowering the energy of the
pseudo-Goldstone excitation.

Besides the dispersion relations (rotons and pseudo-Goldstone modes), an important aspect of the
experimental observations on KCSO is that a large part of the spectral weight of neutron scattering
is distributed over an excitation continuum, rather than in sharp quasi-particle
peaks~\cite{zhu_prl_2024, chen_nc_2026, zhu_npj_2025}.
It has been suggested that the intensity measured experimentally at zero field arises predominantly
from the longitudinal structure factor ${\cal S}^{zz}(\bq, \omega)$ (which corresponds to the
density structure factor in the boson language)~\cite{zhu_npj_2025}.

The strong-coupling $V \gg 1$ of KCSO makes it impossible to derive quantitative conclusions on the
distribution of the spectral weight at arbitrary energies from the one-loop calculations.
However, we can give some qualitative discussions, and present partial results from our analysis in
the small-$V$ region.

Due to the presence of a condensate, the longitudinal structure factor ${\cal S}^{zz}(\bq, \omega)$
receives contributions from states with both even and odd numbers of
excitations~\cite{mourigal_prb_2013, sheng_innov_2025}.
Within a formal small-$V$ expansion, single-magnon modes contribute at order $1/V$ and provide the
leading term, while two-magnon states are suppressed and contribute starting at order $V^{0}$.

The intensity of the single-magnon peaks in the linear-spin wave approximation (renormalized by the
inclusion of a field $\tilde{b}$ to gap the pseudo-Goldstone branch) is, from an explicit
calculation:
\begin{equation} \label{1mag}
\begin{split}
{\cal S}^{zz}(\bq, \omega) & = \sum_{\lambda= \pm 1} \frac{2\pi}{3 V} \lt(1 - \frac{b}{3}\rt)
\frac{(3 + \lambda |f_{\bq}|)}{\omega} \\
& \times \lt(1 - \frac{\lambda {\rm Re}f_{\bq}}{|f_{\bq}|}\rt) \delta(\omega -
\epsilon_{\lambda}(\bk))~.
\end{split}
\end{equation}

The corrections of order $V^{0}$ contribute via self-energy corrections to the single-magnon term,
and via new terms reflecting correlations of the form $\langle \delta m (\delta m)^{2}\rangle$ and
$(\langle \delta m)^{2} (\delta m)^{2} \rangle$.
While it is formally suppressed in the semiclassical expansion, the correlation $\langle (\delta
m)^{2} (\delta m)^{2}\rangle$ is essential at low energies because, as in every 2D superfluid, it
generates a singular contribution $\propto (\omega^{2} - v_{s}^{2} k^{2})^{-1/2}$ near the gapless
points.
This singular contribution arises from the continuum of two low-energy Goldstone modes, and gives
rise to a logarithmically divergent static structure factor, which dominates the correlations
at low momenta~\cite{castellani_prl_1997, kreisel_prb_2014}.
Since this effect is universal in 2D superfluids, it is robust for any values of the coupling $V$
and has validity also in the strong coupling regime.

Other effects connected with low-energy Goldstone fluctuations are expected in the self-energy
corrections.
Because of the presence of gapless modes, the single-magnon dispersion at all momenta $\bk$
coincide with the end point of branch cuts in the spectral function.
The infrared divergence of the self-energy matrix in Sec.~\ref{corrections_to_the_spectrum} can be
traced to the coincidence between branch cuts and single-magnon dispersion; because this degeneracy
is only due to symmetry, it is also expected to be robust in the large-$V$ limit.

Analyzing in more details whether low-energy Goldstone modes play an important role in the continua
observed in KCSO is beyond the scope of the present work.
An indication that low-energy modes may be relevant comes from the fact that the experimental data
do not show any gap between sharp features at the overall bottom of the spectrum and the bottom of
the continuum.

\begin{figure}
\centering
\includegraphics[scale=1]{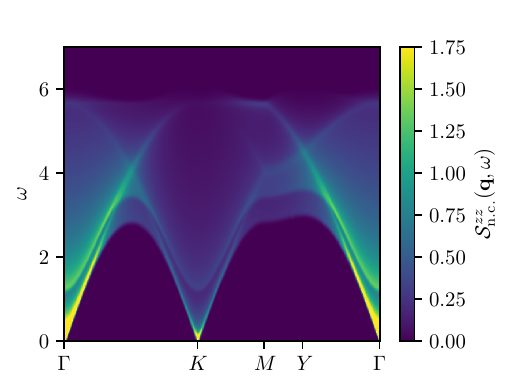}
\caption{\label{2mag} Contribution to ${\cal S}^{zz}(\bq, \omega)$ arising from density-density
correlation of non-condensed bosons, calculated within the linear-spin-wave approximation.
In the calculation we considered $\tilde{b} = 0.03$, corresponding to $V \simeq 0.05$.
This value of $\tilde{b}$ mimicks the gap observed experimentally in KCSO.}
\end{figure}

As a first step towards the description of continuum excitations, we calculated the contribution to
${\cal S}^{zz}$ due to the term $\langle (\delta m)^{2} (\delta m)^{2}\rangle$, which correspond to
density correlations of non-condensed particles.
In the calculation we use the linear approximation, including the renormalized field $\tilde{b}$ to
mimick the pseudo-Goldstone gap (see Appendix~\ref{a_2mag}).
In the small-$V$ limit, this term constitutes one among the leading-order corrections to
linear-spin-wave theory.
The results are shown in Fig.~\ref{2mag}.

At low-energy, near $\Gamma$ and $K$, ${\cal S}^{zz}_{\rm n.c.}(\bq, \omega)$ behaves as  ${\cal
S}^{zz} \propto \lt(\omega^{2} -v_{\rm s}^{2} q^{2} \rt)^{-1/2}$ due to soft magnon fluctuations.
At higher energy, and near the $K$ point, ${\cal S}^{zz}_{\rm n.c.}$ shows a clearly visible edge.
We verified that this edge occurs at $\omega = \epsilon_{1}(\bq) + \epsilon_{\rm g}$, which
corresponds to the threshold for the emission of two pseudo-Goldstone excitations.
At $\omega = \epsilon_{1}(\bq)$, corresponding to the edge for 1 pseudo-Goldstone and
1 Goldstone excitation, the figure does not show instead a visible singularity.
We verified however that the spectral function, in reality, presents jump-like singularities at
both $\omega = \epsilon_{1}(\bq) + \epsilon_{\rm g}$ and $\epsilon_{1}(\bq)$; only the amplitude
of the latter jump is much weaker than that of the former, in the approximations which we
considered and near the $K$ point.

Another notable feature of the distribution of intensity shown in Fig.~\ref{2mag} is that ${\cal
S}_{\rm n.c.}^{zz}(\bq, \omega)$ presents, overall, a stronger weight near the $\Gamma$ point than
near the $K$ point.
Experiments on KCSO, show instead, a stronger intensity near the $K$ point.

The description of experiments on KCSO eventually requires to describe the limit $V \to \infty$ in
which the problem has no small parameter, and all terms are a priori of the same order.
The fact that the distribution of intensity between the $\Gamma$ and the $K$ points cannot be
explained by the ${\cal S}_{\rm n.c.}^{zz}(\bq, \omega)$ could be viewed, however, as a first
tentative indication that the contribution of the ``non-condensed'' particles, is not the dominant
term to explain the continuum observed near the $K$ point.

\section{Summary and conclusions}
\label{summary_and_conclusions}

In summary, we have derived analytical expressions for the self-energy corrections at one-loop
order in a boson model on the honeycomb lattice with nearest-neighbour repulsion.
The model, under some assumptions, is equivalent to the large-$S$ limit of the triangular XXZ
antiferromagnet in the limit of strong anisotropy.

For zero field, the corresponding Hamiltonian presents a one-dimensional family of accidentally
degenerate classical ground states, leading to a pseudo-Goldstone excitation in spin-wave theory.
By calculating the self-energy analytically, we derived an expression for the pseudo-Goldstone gap,
and for the spectrum renormalization at arbitrary momenta for zero magnetic field.
Our analysis is quantitatively controlled in the limit $S \gg 1/\alpha$, and extends to one-loop
order the linear spin wave approximation discussed in Ref.~\cite{kleine_zpb_1992b}.

Infrared divergences in the expansion of the spectrum the calculation introduce an additional
difficulty in comparison with the spin-wave expansion of the Heisenberg model on the triangular
lattice.
In this work, we studied the spectrum using an approximate self-consistent spin-wave calculation
including an energy bandwidth renormalization and a magnetic field renormalization, which has the
effect to cutoff the IR singularities.
Within this framework, we find that the corrections present a complex dependence on $V$ and induce
an (avoided) crossing of magnon branches.

The analyses presented in this work leave open some crucial questions which are left for future
study.
In the limit of moderately small $V$, a natural extension of our analysis consists in a computation
of the spectrum beyond the on-shell approximation.
This generalization is necessary in order to regularize singularities in the spectrum generated by
topological transitions in the decay surface, in analogy with the discussions presented in
Ref.~\cite{chernyshev_prb_2009}.
In the Heisenberg model on the triangular lattice, the analysis of Ref.~\cite{chernyshev_prb_2009}
indicated that the spectrum determined by the solution of ``off-shell'' Dyson equations not only
regularizes singularities, but also produces more a pronounced roton-like minimum than the
first-order $1/S$ expansion.
More generally, further analyses are needed to address the regime of moderate $V$, and the
stability of our results to the inclusion of higher-order nonlinearities and more complex
self-consistent renormalization schemes.

Finally, we discussed qualitatively the structure of the nonlinear spin-wave corrections with the
spectrum observed in KCSO.
Our analysis shows that it is not possible to choose a small value of $V$ reproducing
simultaneously a roton-minimum and a small pseudo-Goldstone gap.
Our analysis however can only be a first qualitative discussion, because our analysis is based on
a small $V$ expansion, while a theory of KCSO requires an understanding of the strong-coupling
region $V \to \infty$.
An analysis beyond that presented here is essential to understand the strong-coupling regime.

A question which remains open, in particular, concerns the magnitude of the pseudo-Goldstone gap.
The semiclassical interpretation of this excitation as a pseudo-Goldstone mode arising from the
classical degeneracy links its understanding to spin-wave theory.
However, the analysis presented here shows that the semiclassical approximation leads to a gap
which grows very rapidly with $V$, overcoming the experimental value already at $V \simeq 0.05$.
This raises the question of how the pseudo-Goldstone gap evolves between weak-coupling and
strong-coupling limits.

\emph{Note}---While we were completing this work we became aware of a preprint,
Ref.~\cite{lin_arxiv_2025}, which discussed the pseudo-Goldstone gap and the $1/S$
corrections to the spectrum in the triangular XXZ model at finite temperatures.

\emph{Acknowledgments}---We are grateful to C. Batista, A. L\"{a}uchli, and O.  Starykh, for
discussions on related topics. This work has been supported by the Swiss National Science Foundation
Grant No. 212082.

\emph{Data availability}---The data that support the findings of this article are openly
available~\cite{mauri_zenodo_nlswt}.

\appendix

\section{Non-interacting theory and bare propagators}
\label{bare_propagators}

To derive the spectrum and the propagator in linear spin wave theory, we expand Eq.~\eqref{H_ef} at
the semiclassical minimum $m^{\alpha}_{i} = m^{\alpha}_{i0} = r_{i} \sqrt{2(1 - b/3)/V}
\delta^{\alpha x}$ and we keep terms of second order in the fluctuations $\delta m^{\alpha}_{i}$.
The quadratic part gives the noninteracting Hamiltonian
\begin{equation}
\begin{split}
H_{0} & = \sum_{\langle i, j \rangle} \lt(\delta m_{i}^{y} \delta m_{i}^{y} -
 \lt(1 - \frac{2}{3} b\rt)\delta m_{i}^{x} \delta m_{j}^{x}  \rt)  \\
& \qquad + \frac{3}{2} \sum_{i}  \delta m_{i}^{\alpha} \delta m_{i}^{\alpha}\\
& = \sum_{\langle i, j \rangle} \lt(\frac{b}{3}\delta^{\alpha \beta} - \lt(1 -
\frac{b}{3}\rt) \sigma_{z}^{\alpha \beta}\rt)\delta m_{i}^{\alpha} \delta m_{j}^{\beta}\\
& \qquad + \frac{3}{2} \sum_{i} \delta m_{i}^{\alpha} \delta m_{i}^{\alpha}~.
\end{split}
\end{equation}

The commutation relations $[\delta m^{\alpha}_{i}, \delta m^{\beta}_{j}] = -\delta_{ij}
\sigma_{y}^{\alpha \beta}$ and the equations of motion $\delta \dot{m}_{i}^{\alpha} = i [H,
\delta m^{\alpha}_{i}]$ imply that the bare time-ordered Green function $G^{\alpha \beta}_{0ij}(t -
t') = -i \langle 0| T \{ \delta m^{\alpha}_{i}(t) \delta m_{j}^{\beta}(t')\} |0 \rangle = -i
\Theta(t-t') \langle 0|\delta m_{i}^{\alpha}(t) \delta m_{j}^{\beta}(t') |0\rangle - i \Theta(t'-t)
\langle 0|\delta m_{j}^{\beta}(t') \delta m_{i}^{\alpha}(t)|0\rangle$, satisfies:
\begin{equation}
\begin{split}
& -i \sigma_{y}^{\alpha \gamma} \frac{{\rm d}}{{\rm d}t} G^{\gamma \beta}_{0ij}(t-t') = \delta_{ij}
\delta^{\alpha \beta} \delta(t-t') + 3 G_{0 ij}^{\alpha \beta}(t-t')\\
& + \sum_{k \in {\rm NN}(i)} \lt(\frac{b}{3} \delta^{\alpha \gamma} - \lt(1 - \frac{b}{3}\rt)
\sigma_{z}^{\alpha \gamma}\rt) G_{0kj}^{\gamma \beta}(t-t')~.
\end{split}
\end{equation}

Here and in the following $\sum_{k \in {\rm NN}(i)}$ stands for a summation over the three sites
which are nearest neighbours of site $i$ on the honeycomb lattice.
The inversion of the equation is performed introducing the Fourier representation
\begin{equation} \label{G_fourier}
\begin{split}
G^{\alpha \beta}_{0ij}(t-t') & = \int_{-\infty}^{\infty} \frac{{\rm d}\omega}{2 \pi} \int_{{\rm BZ}}
\frac{{\rm d}^{2}k}{\Omega} \big[ {\rm e}^{i \bk \cdot (\bx_{i} - \bx_{j}) - i \omega (t-t')} \\
& \times G^{\alpha \beta}_{\tau_{i} \tau_{j}}(\bk, \omega) \big]~,
\end{split}
\end{equation}
where $\tau_{i}$, $\tau_{j}$ are sublattice indices ($\tau_{i} =+1$ if $i$ is in sublattice A and
$\tau_{i} = 2$ if $i$ is in sublattice B).

In momentum space the equation of motion can be written as:
\begin{equation} \label{G_EOM}
\begin{split}
& \delta_{\tau \tau'}\delta^{\alpha \beta} = - \sum_{\tau''} \Big[3  \delta_{\tau
\tau''}\delta^{\alpha \gamma} + \omega \delta_{\tau \tau''}\sigma_{y}^{\alpha \gamma} +
\Big(\frac{b}{3} \delta^{\alpha \gamma} \\
& - \lt(1  - \frac{b}{3}\rt)\sigma_{z}^{\alpha \gamma}\Big) F_{\tau \tau''}(\bk) \Big] G^{\gamma
\beta}_{\tau'' \tau'}(\bk, \omega)~,
\end{split}
\end{equation}
where
\begin{equation}
F(\bk) = \begin{vmatrix}
        0 & f_{\bk}^{*} \\
        f_{\bk} & 0
       \end{vmatrix}~.
\end{equation}

(Summation over all repeated cartesian indices is implicitly assumed everywhere).

Taking into account the pole prescription for the time-ordered Green function, we find that
$G_{0}(\bk, \omega)$ is the inverse of $G_{0}^{-1}(\bk, \omega) = - \lt(3 + (1 + i0^{+})\omega
\sigma_{y} + \lt(\frac{b}{3} - \lt(1 - \frac{b}{3}\rt) \sigma_{z} \rt) F_{\bk}\rt)$.
The $4 \times 4$ matrix $G_{0}^{-1}$ can be written in block form as
\begin{equation}
G_{0}^{-1}(\bk, \omega) =  -\begin{Vmatrix}
 3 I - \lt(1 - \frac{2}{3}b\rt) F_{\bk} & -i (1 + i0^{+}) \omega I\\
 i (1 + i0^{+}) \omega I &  3 I + F_{\bk}
                          \end{Vmatrix}~,
\end{equation}
with $I$ the $2 \times 2$ identity matrix.
The dispersion relations of the magnon modes can be determined equivalently, from the poles of
$G_{0}(\bk, \omega)$ or from the condition $\det G_{0}^{-1}(\bk, \epsilon_{\bk}) = 0$.

For zero magnetic field, the propagator and the spectrum can be calculated immediately since the
matrix $G_{0}^{-1} = -(3 + (1 + i0^{+}) \omega \sigma_{y} - F(\bk) \sigma_{z})$ can be directly
inverted as:
\begin{equation}
G_{0}(\bk, \omega) = \frac{3 - \omega \sigma_{y} + F(\bk) \sigma_{z}}{\omega^{2} - (9 -
|f_{\bk}|^{2})+ i0^{+}}~.
\end{equation}
The magnon poles have energy $\epsilon_{0}(\bk) = \sqrt{9 - |f_{\bk}|^{2}}$, which is the
noninteracting spin-wave dispersion at zero field (see Eq.~\eqref{LSWT_ap_res} in the main
text).

At finite field $0 < b < 3$, instead, the algebra becomes longer.
However, even in this case, a significant simplification occurs due to the fact that the Green
function can be diagonalized simultaneously with $F(\bk)$.
This special property be traced to the fact that the quadratic (linear-spin-wave) Hamiltonian has
only nearest-neighbour interactions.

After diagonalizing $F(\bk)$, the $4\times 4$ matrix Green function breaks into two sectors,
corresponding to the two eigenvalues $\pm |f_{\bk}|$ of the matrix $F(\bk)$, and thus we can
invert the matrix considering the two sectors independently.

Within the sector in which $F(\bk) = \lambda |f_{\bk}|$ (with $\lambda = \pm 1$), the Green function
requires to invert the $2 \times 2$ matrix:
\begin{equation}
\begin{split}
-G_{0\lambda}^{-1}(\bk, \omega) & = 3 + (1 + i0^{+})\omega \sigma_{y} \\
& + \lt(\frac{b}{3} - \lt(1 - \frac{b}{3}\rt) \sigma_{z}\rt) \lambda |f_{\bk}|~.
\end{split}
\end{equation}

The inversion is straightforward and gives Eq.~\eqref{G0_propagator} in the main text.

After $G_{0}$ is determined within the subspaces at fixed $\lambda$, the Green function in the
original sublattice basis can be expressed using the unitary transformation $U(\bk)$ which
diagonalizes $F(\bk)$.
This gives Eq.~\eqref{G_unitary} in the main text.

We finally discuss the real-time correlations of the plane wave operators $\delta
m_{\tau}^{\alpha}(\bk, t) = \sqrt{3/N} \sum_{i \in {\rm subl.} \tau} \delta m_{i}^{\alpha}(t) {\rm
e}^{-i \bk \cdot \bx_{i}}$, which are given by the Fourier transform of $G$ over
frequency (but not over momentum).
We find:
\begin{equation}\label{G_time}
\begin{split}
& G_{0\tau \tau'}^{\alpha \beta}(\bk, t-t') = -i \langle 0|T \{\delta m_{\tau}^{\alpha}(\bk, t)
\delta
m_{\tau'}^{+\beta}(\bk, t') \}|0 \rangle \\
& = \int_{-\infty}^{\infty} \frac{{\rm d}\omega}{2 \pi}~G_{0\tau \tau'}^{\alpha
\beta}(\bk, \omega) {\rm e}^{-i\omega (t-t')}\\
& = \sum_{\lambda = \pm 1} \frac{-i}{2\epsilon_{\lambda}(\bk)} U_{\tau \lambda}(\bk) U^{*}_{\tau'
\lambda}(\bk) \\
& \times \bigg[\Theta(t-t') \big( N^{\alpha \beta}_{\lambda}(\bk)  - \epsilon_{\lambda}(\bk)
\sigma_{y}^{\alpha \beta}\big) {\rm e}^{-i \epsilon_{\lambda}(\bk) (t-t')}\\
& + \Theta(t'-t) \big(N_{\lambda}^{\alpha \beta}(\bk) + \epsilon_{\lambda}(\bk) \sigma_{y}^{\alpha
\beta}\big) {\rm e}^{-i \epsilon_{\lambda}(\bk)(t'-t)} \bigg]~,
\end{split}
\end{equation}
where
\begin{equation}
N^{\alpha \beta}_{\lambda}(\bk) = \lt(3 + \frac{b}{3} \lambda |f_{\bk}|\rt) \delta^{\alpha \beta} +
\lt(1 - \frac{b}{3}\rt) \lambda |f_{\bk}|\sigma_{z}^{\alpha \beta}~.
\end{equation}

\subsection*{Equal-time second-order correlations}
\label{2nd_order_correlations}

From the equal-time limit of Eq.~\eqref{G_time}, we obtain the static correlations
\begin{equation}
\begin{split}
& \langle \delta m^{\alpha}_{\tau}(\bk) \delta m^{*\beta}_{\tau'}(\bk) \rangle = i G^{\alpha
\beta}_{\tau \tau'}(\bk, 0^{+}) \\
& = -\frac{1}{2}\delta_{\tau \tau'} \sigma_{y}^{\alpha \beta} + \sum_{\lambda = \pm 1}
P_{\lambda \tau \tau'}(\bk) \frac{N_{\lambda}^{\alpha \beta}(\bk)}{2 \epsilon_{\lambda}(\bk)}~.
\end{split}
\end{equation}

Transforming to real space we find that the static single-site correlation function is
\begin{equation}
\langle \delta m^{\alpha}_{i} \delta m^{\beta}_{i} \rangle_{0} = - \frac{1}{2} \sigma^{\alpha
\beta}_{y} + \sum_{\lambda = \pm 1} \int \frac{{\rm d}^{2}k}{\Omega} \frac{N^{\alpha
\beta}_{\lambda}(\bk)}{4 \epsilon_{\lambda}(\bk)}~.
\end{equation}

In particular, the average $c_{1} = M^{\alpha \beta} \langle \delta m^{\alpha}_{i} \delta
m^{\beta}_{i} \rangle /2 $ is:
\begin{equation} \label{density_normal}
\frac{1}{2} M^{\alpha \beta} \langle \delta m^{\alpha}_{i} \delta m^{\beta}_{i}\rangle_{0}
= - \frac{1}{2} + \sum_{\lambda = \pm 1} \int \frac{{\rm d}^{2}k}{\Omega} \frac{\lt(3 + \frac{b}{3}
\lambda|f_{\bk}|\rt)}{4 \epsilon_{\lambda}(\bk)}~.
\end{equation}

The correlation in Eq.~\eqref{density_normal} is related to the average number of bosons per site
in the linear spin wave approximation: $\langle n_{i} \rangle = |\varphi|^{2} + M^{\alpha \beta}
\langle \delta m^{\alpha}_{i} \delta m^{\beta}_{i}\rangle/2$.

Since it will be used in the calculation of the order parameter renormalization, we also calculate
the nearest-neighbour correlation.
This can be calculated from the Fourier-space correlation~\eqref{G_time} at $t = 0^{+}$.
Assuming that $i$ is in sublattice A and $j$ in sublattice B we find:
\begin{equation}
\langle \delta m^{\alpha}_{i} \delta m^{\beta}_{j} \rangle_{0} = \sum_{\lambda = \pm 1} \int
\frac{{\rm d}^{2}k}{\Omega} \frac{\lambda f^{*}_{\bk} N_{\lambda}^{\alpha \beta}(\bk)}{4 |f_{\bk}|
\epsilon_{\lambda}(\bk)} {\rm e}^{i \bk \cdot (\bx_{i} - \bx_{j})}~.
\end{equation}

The symmetries of the lattice imply that the correlation does not depend on the direction on the
nearest-neighbour bond; as a result, the expression can be simplified, without changing the result,
by ``averaging'' over the three direction of the bond.
As a result we find:
\begin{equation}
\langle \delta m^{\alpha}_{i} \delta m^{\beta}_{j} \rangle_{0} = \frac{1}{12} \sum_{\lambda = \pm 1}
\int \frac{{\rm
d}^{2}k}{\Omega} \frac{\lambda |f_{\bk}| N_{\lambda}^{\alpha \beta}(\bk)}{\epsilon_{\lambda}(\bk)}
\end{equation}

In particular, for the correlations in the $x$ direction we find:
\begin{equation}
c_{2} = \langle \delta m^{x}_{i} \delta m^{x}_{j} \rangle_{0} = \frac{1}{12} \sum_{\lambda = \pm
1} \int_{{\rm BZ}} \frac{{\rm d}^{2}k}{\Omega} \frac{(3 + \lambda |f_{\bk}|) \lambda
|f_{\bk}|}{\epsilon_{\lambda}(\bk)}~,
\end{equation}
and for the correlations in the $y$ direction
\begin{equation}
\begin{split}
c_{3} & = \langle \delta m^{y}_{i} \delta m^{y}_{j} \rangle_{0} \\
& = \frac{1}{12} \sum_{\lambda = \pm 1}
\int \frac{{\rm d}^{2}k}{\Omega} \frac{\lt(3  - \lt(1 - \frac{2}{3} b\rt) \lambda
|f_{\bk}|\rt)\lambda |f_{\bk}|}{\epsilon_{\lambda}(\bk)}~.
\end{split}
\end{equation}

\section{Correction to the order parameter}
\label{order_parameter}

At the classical level, the order parameter is $m_{i0}^{\alpha} = \sqrt{2} \varphi r_{i}$
with $r_{i} = \pm 1$ on the two sublattices, and with $\varphi = \varphi_{\rm cl} =
\sqrt{(1-b/3)/V}$.
At one-loop order, the order parameter $\varphi$ receives corrections from magnon fluctuations.

A possible way to compute $\varphi$ consists in replacing in the Hamiltonian~\eqref{H_ef_1}
$m_{i}^{\alpha} \to \bar{m}_{i}^{\alpha} + \delta m^{\alpha}_{i}$.
After this substitution, the Hamiltonian presents terms linear, quadratic, cubic, and quartic in
$\delta m^{\alpha}_{i}$.
The order parameter can then be determined by requiring that the linear part in $\delta
m^{\alpha}_{i}$ is balanced by an equal and opposite ``force'' coming from the fluctuations.
This force in particular be derived from the cubic terms in $\delta m$ by decoupling $\delta
m^{\alpha}_{i} \delta m^{\beta}_{j} \delta m^{\gamma}_{k} \simeq \delta m^{\alpha}_{i} \langle
\delta m^{\beta}_{j} \delta m^{\gamma}_{k} \rangle + \delta m^{\beta}_{j} \langle \delta
m^{\alpha}_{i} \delta m^{\gamma}_{k} \rangle + \delta m^{\gamma}_{k} \langle \delta m^{\alpha}_{i}
\delta m^{\beta}_{j} \rangle$.

A more systematic derivation however can be given using the equation of motion
$\dot{m}^{\alpha}_{i} = i [H, m^{\alpha}_{i}]$, which, taking the ground-state average, implies
$\langle \dot{m}^{\alpha}_{i} \rangle = i \langle [H, m^{\alpha}_{i}] \rangle = 0$.
From this relation it follows that the order parameter satisfies the exact relation:
\begin{equation} \label{eom}
\sum_{j \in {\rm NN}(i)} \lt( \bar{m}^{\alpha}_{j} + \frac{V}{2} M^{\beta \gamma}  \lt
\langle m^{\alpha}_{i} m^{\beta}_{j} m^{\gamma}_{j} \rt \rangle \rt) + b
\bar{m}_{i}^{\alpha} = 0~.
\end{equation}

In the classical limit, when $\langle m^{\alpha}_{i} m^{\beta}_{j} m^{\gamma}_{j}\rangle \to
\bar{m}^{\alpha}_{i} \bar{m}^{\beta}_{j} \bar{m}^{\gamma}_{j}$, Eq.~\eqref{eom} gives the classical
order parameter $m_{i0}^{\alpha}$.
At one-loop order, instead, the cubic average has to be replaced with
\begin{equation} \label{cubic_average}
\begin{split}
\langle m^{\alpha}_{i} m^{\beta}_{j} m^{\gamma}_{j} \rangle & = \bar{m}^{\alpha}_{i}
\bar{m}^{\beta}_{j} \bar{m}^{\gamma}_{j} + \bar{m}_{i}^{\alpha} \langle \delta
m_{j}^{\beta} \delta m_{j}^{\gamma} \rangle_{0}  \\
& + \bar{m}_{j}^{\beta} \langle \delta m_{i}^{\alpha} \delta m_{j}^{\gamma} \rangle_{0}
+  \bar{m}_{j}^{\gamma} \langle \delta m_{i}^{\alpha} \delta m_{j}^{\beta} \rangle_{0}~,
\end{split}
\end{equation}
where $\langle \delta m^{\alpha}_{i} \delta m^{\beta}_{j} \rangle_{0}$ are equal-time correlations
of the fluctuations $\delta m^{\alpha}_{i}$ calculated in the non-interacting model.

Solving Eq.~\eqref{eom} together with Eq.~\eqref{cubic_average} we obtain:
\begin{equation}
\varphi^{2} =\frac{1}{V}\lt(1 - \frac{b}{3}\rt) + \lt \langle \delta m^{x}_{i} \delta m^{x} _{j}
\rt \rangle_{0} - \frac{1}{2} M^{\alpha \beta} \lt \langle \delta m^{\alpha}_{i} \delta
m^{\beta}_{i} \rt \rangle_{0}~,
\end{equation}
where $i$ and $j$ are any pair of nearest-neighbour sites on the lattice.

Using $\varphi_{\rm cl}^{2} = (1 - b/3)/V$ and the results of Sec.~\ref{2nd_order_correlations} we
find:
\begin{equation} \label{integral_order_parameter}
\begin{split}
& \varphi^{2} - \varphi_{\rm cl}^{2} =  \frac{1}{2}  -\frac{1}{12} \sum_{\lambda = \pm 1} \int_{\rm
BZ} \frac{{\rm d}^{2}k}{\Omega} \frac{9 - (3 - b + \lambda |f_{\bk}|) \lambda
|f_{\bk}|}{\epsilon_{\lambda}(\bk)} \\
& \qquad + O(V)~.
\end{split}
\end{equation}

\begin{figure}[h]
\centering
\includegraphics[scale=1]{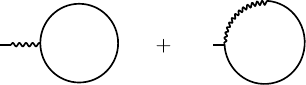}
\caption{\label{tadpoles} Tadpole diagrams determining the correction to the order parameter.}
\end{figure}

Diagrammatically, the integral in Eq.~\eqref{integral_order_parameter} corresponds to the tadpoles
in Fig.~\ref{tadpoles}.

\section{Calculation of the self-energy}
\label{ap_self_energy}

\subsection{Tadpole term $\Sigma_{a}$}

The contribution of the tadpole graph~\ref{diagrams}a can be calculated in the one-loop
approximation by calculating the variation of $\pa^{2} H/(\pa m_{i}^{\alpha} \pa
m_{j}^{\beta})|_{\bar{m}_{i}^{\alpha}}$ due to the shift of the order parameter
$\bar{m}_{i}^{\alpha} - m_{i0}^{\alpha}$.
Calculating the second derivative explicitly from Eq.~\eqref{H_ef_1} and expanding about the
average configuration $\bar{m}_{i}^{\alpha} = \sqrt{2} \varphi r_{i}$ we obtain:
\begin{equation}
\begin{split}
\frac{\pa^{2}H}{\pa m_{i}^{\alpha} \pa m_{j}^{\beta}}\bigg|_{\bar{m}_{i}^{\alpha}} & =
\big((1 - V \varphi^{2}) t_{ij} + (3 V \varphi^{2} + b)\delta_{ij} \big) \delta^{\alpha \beta} \\
& \quad  - V \varphi^{2}  t_{ij} \sigma_{z}^{\alpha \beta}~.
\end{split}
\end{equation}

Replacing $\varphi^{2}$ with its classical value $\varphi_{\rm cl}^{2} = (1 - b/3)/V$ gives the
magnon Hamiltonian, which determines the linear spin wave spectrum.
The correction
\begin{equation}
\begin{split}
& \frac{\pa^{2}H}{\pa m_{i}^{\alpha} \pa m_{j}^{\beta}}\bigg|_{\bar{m}_{i}^{\alpha}} -
\frac{\pa^{2}H}{\pa m_{i}^{\alpha} \pa m_{j}^{\beta}}\bigg|_{m_{i0}^{\alpha}}  \\
& = V (\varphi^{2} - \varphi_{\rm cl}^{2}) ((3 \delta_{ij} -t_{ij})  \delta^{\alpha \beta} - t_{ij}
\sigma_{z}^{\alpha \beta})~,
\end{split}
\end{equation}
gives the contribution $\Sigma_{\rm a}$ to the self-energy.

In momentum space, we find:
\begin{equation}
\begin{split}
\Sigma_{{\rm a}\tau \tau'}^{\alpha \beta}(\bk, \omega) & = V(\varphi^{2} - \varphi^{2}_{\rm cl}) ((3
\delta_{\tau \tau'} - F_{\tau \tau'}(\bk)) \delta^{\alpha \beta} \\
& \qquad - F_{\tau \tau'}(\bk) \sigma_{z}^{\alpha \beta})~.
\end{split}
\end{equation}
Transforming from the sublattice to the branch basis we can replace $F(\bk) = \lambda |f_{\bk}|$.
We then obtain Eq.~\eqref{sigma_a} in the main text.

\subsection{Hartree-Fock diagram $\Sigma_{b}$}

The second self-energy graph, $\Sigma_{\rm b}$, can be calculated conveniently in real space and
involves the same static averages which were calculated in Sec.~\ref{2nd_order_correlations}.
Explicitly we find
\begin{equation} \label{sigma_b}
\begin{split}
\Sigma_{{\rm b} ij}^{\alpha \beta}(t-t') & = \bigg[ \frac{3V}{2}   \delta_{ij} \delta^{\alpha
\beta} M^{\gamma \delta} \langle \delta m_{k}^{\gamma} \delta m_{k}^{\delta}\rangle_{0} \\
& + V t_{ij} \langle \delta m_{k}^{\alpha} \delta m_{l}^{\beta} \rangle_{0} \bigg] \delta(t-t')
\end{split}
\end{equation}
where $k$ and $l$ are any pair of nearest-neighbour sites on the honeycomb lattice and $t_{ij}$ is
the adjacency matrix ($t_{ij} = 1$ if $i$, $j$ are nearest-neighbours and $t_{ij} = 0$ otherwise).
The terms in the first and in the second line in Eq.~\eqref{sigma_b} correspond, respectively, to
the Hartree- and Fock-like diagrams in Eq.~\ref{HF} of the main text.

Using $\langle \delta m_{k}^{\alpha} \delta m_{l}^{\beta} \rangle = (c_{2} + c_{3})\delta^{\alpha
\beta}/2 + (c_{2} - c_{3}) \sigma_{z}^{\alpha \beta}/2$, and transforming to momentum space we find:
\begin{equation}
\begin{split}
\Sigma_{{\rm b} ij}^{\alpha \beta}(\bk, \omega) & = V\Big(3 c_{1} \delta_{\tau \tau'}
\delta^{\alpha \beta} + \frac{1}{2} F_{\tau \tau'}(\bk) \big((c_{2} + c_{3})\delta^{\alpha \beta}
\\
& \qquad + (c_{2} - c_{3}) \sigma_{z}^{\alpha \beta} \big)\Big)~.
\end{split}
\end{equation}

In the branch basis this gives Eq.~\eqref{HF}.

\subsection{Frequency-dependent self-energy term $\Sigma_{\rm c}$}
\label{ap_sigma_c}

The last term $\Sigma_{\rm c}$ reads can be expressed in real space as:
\begin{equation}
\begin{split}
\Sigma_{{\rm c} i j}^{\alpha \beta}(t-t') & = -\frac{i}{2} \sum_{l, m, l',
m'} \big[\Gamma^{\alpha \lambda \mu}_{i l m} \Gamma^{\nu \rho \beta}_{l' m' j} G_{0 l
l'}^{\lambda \nu}(t - t') \\
& \times G_{0 m m'}^{\mu \rho}(t - t') \big]~.
\end{split}
\end{equation}
where $\Gamma$ is the three-particle vertex in real space.
The vertex $\Gamma$ is defined via the cubic part in $\delta m$ of the Hamiltonian,
expanded at the classical minimum $m = m_{0}$.
It can be calculated as:
\begin{equation}
\begin{split}
\Gamma^{\alpha \beta \gamma}_{ijk} & = -i \frac{\pa^{3} H}{\pa m^{\alpha}_{i} \pa
m^{\beta}_{j} \pa m^{\gamma}_{k}}\bigg|_{m = m_{0}} \\
& = -i V \big(\delta_{ij} t_{jk} \delta^{\alpha \beta} m_{k0}^{\gamma} + \delta_{jk}
t_{ki} \delta^{\beta \gamma} m_{i0}^{\alpha} \\
& \qquad + \delta_{ki} t_{ij} \delta^{\gamma \alpha} m_{j0}^{\beta} \big)~,
\end{split}
\end{equation}
where $m_{i0}^{\alpha} = \sqrt{2(1 - b/3)/V} r_{i} \delta^{\alpha 1}$ is the in-plane
magnetization of site $i$ in the classical limit.

In order to transform into momentum space we introduce the Fourier-space vertex
$\tilde{\Gamma}(\bk_{1}, \bk_{2}, \bk_{3})$ via the relations:
\begin{equation}\label{fourier_vertex}
\begin{split}
& \sum_{i \in {\rm subl.}\tau} \sum_{j \in
{\rm subl.} \tau'} \sum_{k \in {\rm subl.} \tau''} \Gamma_{ijk}^{\alpha \beta
\gamma} {\rm e}^{-i (\bk_{1} \cdot \bx_{i} + \bk_{2} \cdot \bx_{j} +
\bk_{3} \cdot \bx_{k})} \\
& = \sum_{\bG} \Omega \delta(\bk_{1} + \bk_{2} + \bk_{3}- \bG) \tilde{\Gamma}^{\alpha
\beta \gamma}_{\tau \tau' \tau''}(\bk_{1}, \bk_{2}, \bk_{3})~,
\end{split}
\end{equation}
\begin{equation}
\begin{split}
& \tilde{\Gamma}_{\tau \tau' \tau''}^{\alpha \beta \gamma}(\bk_{1}, \bk_{2}, \bk_{3}) =
-i V \big( \\
& \qquad {\rm e}^{-i \bG \cdot \ba_{\tau}}\delta_{\tau \tau'} F_{\tau''
\tau'}(\bk_{3}) \delta^{\alpha \beta} m_{\tau''0}^{\gamma}\\
& \qquad + {\rm e}^{-i \bG \cdot \ba_{\tau'}} \delta_{\tau' \tau''}  F_{\tau
\tau''}(\bk_{1}) \delta^{\beta \gamma} m_{\tau0}^{\alpha}\\
& \qquad + {\rm e}^{-i \bG \cdot \ba_{\tau''}}\delta_{\tau'' \tau} F_{\tau'
\tau}(\bk_{2}) \delta^{\gamma \alpha} m_{\tau' 0}^{\beta}\big)~,\\
& \bG = \bk_{1} + \bk_{2} + \bk_{3}~.
\end{split}
\end{equation}

In these expressions, $\bG$ runs over the reciprocal lattice vectors of the
honeycomb lattice ($\bG = n_{1} \bG_{1} + n_{2} \bG_{2}$ with $n_{1}$, $n_{2}$ integer,
$\bG_{1}= \lt(\frac{4\pi}{3}, 0\rt)$, $\bG_{2} = \big(-\frac{2 \pi}{3}, \frac{2 \sqrt{3}
\pi}{3}\big)$).
The delta function expresses momentum conservation (up to a reciprocal lattice vector).
In addition, with a natural notation, we denoted as $m_{\tau 0}^{\alpha}$ the classical
moment in sublattice $\tau$, $m_{\tau 0}^{\alpha} = \big(\sqrt{2 (1 - b/3)/V}
\big) r_{\tau} \delta^{\alpha 1}$ with $r_{\tau} = 1$ in sublattice A and $r_{\tau} = -1$
in sublattice B.
The vectors $\ba_{\tau}$ are the coordinates of any two atoms in the two sublattices.
The simplest choice is $\ba_{1} = (0, 0)$ and $\ba_{2} = (-1, 0)$.
Any other choice of the vectors $\ba_{1}$ and $\ba_{2}$ with $\ba_{1}$ in sublattice A and
$\ba_{2}$ in sublattice B gives the same result.
In fact, since $\bG$ is a reciprocal vector, the phase factors ${\rm e}^{-i \bG\cdot \bx_{i}}$ are
the same for all $i$ within the same sublattice.

We also note that when $\bk_{1} + \bk_{2} + \bk_{3} = \bG$ is a reciprocal lattice vector then:
\begin{equation} \label{fourier_vertex_part}
\begin{split}
& \sum_{i \in {\rm subl.} \tau} \sum_{j \in {\rm subl.} \tau'} \Gamma_{ijk}^{\alpha \beta \gamma}
{\rm e}^{-i (\bk_{1} \cdot \bx_{i} + \bk_{2} \cdot \bx_{j} + \bk_{3} \cdot \bk_{k})} \\
& \qquad = \tilde{\Gamma}^{\alpha \beta \gamma}_{\tau \tau' \tau_{k}}(\bk_{1}, \bk_{2}, \bk_{3})~.
\end{split}
\end{equation}

Using Eqs.~\eqref{fourier_vertex},~\eqref{fourier_vertex_part},~\eqref{G_fourier},
and~\eqref{G_time} we find after a Fourier transformation the self-energy $\Sigma_{{\rm c}}$ in
momentum space:
\begin{equation}
\begin{split}
& \Sigma_{{\rm c} \tau \tau'}^{\alpha \beta}(\bk, \omega) = -\frac{1}{8} \sum_{\lambda_{1},
\lambda_{2}} \sum_{\tau_{1}, \tau_{2}, \tau'_{1}, \tau'_{2}} \sum_{\bG} \Bigg\{ \\
& \int_{{\rm BZ}}  \frac{{\rm d}^{2}k_{1}}{\Omega}  \int_{{\rm BZ}} \frac{{\rm d}^{2}k_{2}}{\Omega}
\Bigg[ \frac{\big( \Omega \delta(\bk -\bk_{1} - \bk_{2} - \bG)
\big)}{\epsilon_{\lambda_{1}}(\bk_{1}) \epsilon_{\lambda_{2}}(\bk_{2})} \\
& \times \tilde{\Gamma}^{\alpha \lambda \mu}_{\tau \tau_{1} \tau_{2}}(\bk, -\bk_{1},
-\bk_{2})\tilde{\Gamma}_{\tau_{1}' \tau_{2}' \tau'}^{\nu \rho \beta}(\bk_{1}, \bk_{2},
-\bk)\\
& \times U_{\tau_{1} \lambda_{1}}(\bk_{1}) U^{*}_{\tau'_{1} \lambda_{1}}(\bk_{1}) U_{\tau_{2}
\lambda_{2}}(\bk_{2}) U^{*}_{\tau'_{2} \lambda_{2}}(\bk_{2}) \\
& \times \bigg[\frac{(N_{\lambda_{1}}^{\lambda \nu}(\bk_{1}) - \epsilon_{\lambda_{1}}(\bk_{1})
\sigma_{y}^{\lambda \nu}) (N_{\lambda_{2}}^{\mu \rho}(\bk_{2}) - \epsilon_{\lambda_{2}}(\bk_{2})
\sigma_{y}^{\mu \rho})}{\omega - \epsilon_{\lambda_{1}}(\bk_{1}) -
\epsilon_{\lambda_{2}}(\bk_{2}) + i0^{+}}\\
& - \frac{(N_{\lambda_{1}}^{\lambda \nu}(\bk_{1}) + \epsilon_{\lambda_{1}}(\bk_{1})
\sigma_{y}^{\lambda \nu}) (N_{\lambda_{2}}^{\mu \rho}(\bk_{2}) + \epsilon_{\lambda_{2}}(\bk_{2})
\sigma_{y}^{\mu \rho})}{\omega + \epsilon_{\lambda_{1}}(\bk_{1}) + \epsilon_{\lambda_{2}}(\bk_{2})
- i0^{+}}\Bigg] \Bigg] \Bigg\}~.
\end{split}
\end{equation}

The expression can be simplified further by contracting the sums over $\tau_{1}$, $\tau_{2}$,
$\tau'_{1}$, $\tau_{2}'$, thus absorbing the unitary matrices $U_{\tau \lambda}(\bk)$ into the
vertex functions (i.e.~doing the calculation in the branch basis).

We thus introduce the vertex:
\begin{equation}\label{Lambda_vertex_1}
\begin{split}
& \Lambda^{\alpha \beta \gamma}_{\lambda \lambda' \lambda''}(\bk_{1}, \bk_{2}, \bk_{3}) =
\sum_{\tau, \tau', \tau''} \big[U^{*}_{\tau \lambda}(\bk_{1}) U^{*}_{\tau' \lambda'}(\bk_{2}) \\
& \qquad \qquad \times U^{*}_{\tau'' \lambda''}(\bk_{3}) \tilde{\Gamma}^{\alpha \beta \gamma}_{\tau
\tau' \tau''}(\bk_{1}, \bk_{2}, \bk_{3}) \big] ~.
\end{split}
\end{equation}

Using the matrix elements of $U$ in Eq.~\eqref{U_mat} we find that the vertex has the simple form:
\begin{equation}
\begin{split}
& \Lambda^{\alpha \beta \gamma}_{\lambda \lambda' \lambda''}(\bk_{1}, \bk_{2}, \bk_{3}) = -\frac{i
V}{\sqrt{8 |f_{\bk_{1}} f_{\bk_{2}} f_{\bk_{3}}|}} \\
&  \times \big({\rm e}^{-i \bG \cdot \ba_{2}}
\lambda \lambda' \lambda'' \sqrt{f^{*}_{\bk_{1}}} \sqrt{f^{*}_{\bk_{2}}} \sqrt{f^{*}_{\bk_{3}}} -
\sqrt{f_{\bk_{1}}} \sqrt{f_{\bk_{2}}} \sqrt{f_{\bk_{3}}}\big)\\
& \times \big(\lambda |f_{\bk_{1}}| m_{0}^{\alpha} \delta^{\beta \gamma}  + \lambda'
|f_{\bk_{2}}| m_{0}^{\beta} \delta^{\gamma \alpha} +  \lambda'' |f_{\bk_{3}}| m_{0}^{\gamma}
\delta^{\alpha \beta} \big)~,
\end{split}
\end{equation}
where $m_{0}^{\alpha}$ is the magnetization on sublattice A, $m_{0}^{\alpha} = \sqrt{2 (1 -b/3)/V}
\delta^{\alpha 1}$, and $\bG = \bk_{1} + \bk_{2} + \bk_{3}$.

Using that $U(-\bk) = U^{*}(\bk)$, the self-energy in this representation can be written as:
\begin{equation}
\begin{split}
& \Sigma_{{\rm c} \lambda \lambda'}^{\alpha \beta}(\bk, \omega) = -\frac{1}{8}  \sum_{\lambda_{1},
\lambda_{2}} \sum_{\bG}  \Bigg\{ \\
& \times \int_{{\rm BZ}}  \frac{{\rm d}^{2}k_{1}}{\Omega}  \int_{{\rm BZ}} \frac{{\rm
d}^{2}k_{2}}{\Omega}
\Bigg[ \frac{\big( \Omega \delta(\bk -\bk_{1} - \bk_{2} - \bG)
\big)}{\epsilon_{\lambda_{1}}(\bk_{1}) \epsilon_{\lambda_{2}}(\bk_{2})} \\
& \times \Lambda^{\alpha \lambda \mu}_{\lambda \lambda_{1} \lambda_{2}}(\bk, -\bk_{1},
-\bk_{2})\Lambda_{\lambda_{1} \lambda_{2} \lambda'}^{\nu \rho \beta}(\bk_{1}, \bk_{2}, -\bk)\\
& \times \bigg[\frac{(N_{\lambda_{1}}^{\lambda \nu}(\bk_{1}) - \epsilon_{\lambda_{1}}(\bk_{1})
\sigma_{y}^{\lambda \nu}) (N_{\lambda_{2}}^{\mu \rho}(\bk_{2}) - \epsilon_{\lambda_{2}}(\bk_{2})
\sigma_{y}^{\mu \rho})}{\omega - \epsilon_{\lambda_{1}}(\bk_{1}) -
\epsilon_{\lambda_{2}}(\bk_{2}) + i0^{+}}\\
& - \frac{(N_{\lambda_{1}}^{\lambda \nu}(\bk_{1}) + \epsilon_{\lambda_{1}}(\bk_{1})
\sigma_{y}^{\lambda \nu}) (N_{\lambda_{2}}^{\mu \rho}(\bk_{2}) + \epsilon_{\lambda_{2}}(\bk_{2})
\sigma_{y}^{\mu \rho})}{\omega + \epsilon_{\lambda_{1}}(\bk_{1}) + \epsilon_{\lambda_{2}}(\bk_{2})
- i0^{+}}\Bigg] \Bigg] \Bigg\}~.
\end{split}
\end{equation}

In Eq.~\eqref{sigma_c_BZ}, the integration momenta $\bk_{1}$, $\bk_{2}$ are both within the first
Brillouin zone, and the delta function imposes the momentum conservation modulo a reciprocal
vector: $\bk = \bk_{1}+\bk_{2}+ \bG$.
The momentum integral however can be expressed equivalently by identifying $\bk_{2} + \bG \to
\bk_{2}$.
With this change of variables $\sum_{\bG} \int \frac{{\rm d}^{2}k_{2}}{\Omega} = \int \frac{{\rm
d}^{2}k_{2}}{\Omega}$.
Thus the $\bk_{2}$ integration extends to the full two-dimensional plane.
In addition, the delta function becomes $\delta(\bk - \bk_{1} - \bk_{2} - \bG) \to \delta(\bk -
\bk_{1} - \bk_{2})$, implying the constraint $\bk_{2} = \bk - \bk_{1}$ (without a reciprocal
umklapp vector).

It can be checked that the integrand function is invariant under the change of variables $\bk_{2}
\to \bk_{2} + \bg$.
In fact, since $f_{\bk + \bg} = {\rm e}^{-i \bg \cdot \ba_{2}} f_{\bk}$, the absolute value
$|f_{\bk_{2} + \bg}| = |f_{\bk_{2}}|$ is invariant.
This in turn, implies the invariance of $\epsilon_{\lambda_{2}}(\bk_{2}) =
\epsilon_{\lambda_{2}}(\bk_{2} + \bg)$ and $N(\bk_{2}) = N(\bk_{2} + \bg)$.
The vertex $\Lambda$ instead changes by a phase factor under the translation $\bk_{2} \to \bk_{2} +
\bg$.
However, the phase factor cancels when taking the product of the two vertex function in
Eq.~\eqref{sigma_c_BZ}.

As a result, $\Sigma_{\rm c}$ can be written equivalently as:
\begin{equation} \label{sigma_c_BZ}
\begin{split}
& \Sigma_{{\rm c} \tau \tau'}^{\alpha \beta}(\bk, \omega) = -\frac{1}{8} \sum_{\lambda_{1},
\lambda_{2}} \Bigg\{ \\
& \times \int_{{\rm BZ}}  \frac{{\rm d}^{2}k_{1}}{\Omega}
\Bigg[ \frac{\Lambda^{\alpha \lambda \mu}_{\lambda \lambda_{1} \lambda_{2}}(\bk, -\bk_{1},
-\bk_{2})\Lambda_{\lambda_{1} \lambda_{2} \lambda'}^{\nu \rho \beta}(\bk_{1}, \bk_{2},
-\bk)}{\epsilon_{\lambda_{1}}(\bk_{1}) \epsilon_{\lambda_{2}}(\bk_{2})} \\
& \times \bigg[\frac{(N_{\lambda_{1}}^{\lambda \nu}(\bk_{1}) - \epsilon_{\lambda_{1}}(\bk_{1})
\sigma_{y}^{\lambda \nu}) (N_{\lambda_{2}}^{\mu \rho}(\bk_{2}) - \epsilon_{\lambda_{2}}(\bk_{2})
\sigma_{y}^{\mu \rho})}{\omega - \epsilon_{\lambda_{1}}(\bk_{1}) -
\epsilon_{\lambda_{2}}(\bk_{2}) + i0^{+}}\\
& - \frac{(N_{\lambda_{1}}^{\lambda \nu}(\bk_{1}) + \epsilon_{\lambda_{1}}(\bk_{1})
\sigma_{y}^{\lambda \nu}) (N_{\lambda_{2}}^{\mu \rho}(\bk_{2}) + \epsilon_{\lambda_{2}}(\bk_{2})
\sigma_{y}^{\mu \rho})}{\omega + \epsilon_{\lambda_{1}}(\bk_{1}) + \epsilon_{\lambda_{2}}(\bk_{2})
- i0^{+}}\Bigg] \Bigg] \Bigg\}~,
\end{split}
\end{equation}
where $\bk_{2} = \bk - \bk_{1}$.
We note in addition that in this representation the vertices $\Lambda(\bk, -\bk_{1}, -\bk_{2})$ and
$\Lambda(-\bk, \bk_{1}, \bk_{2})$ carry zero total momentum $\bG = \bk - \bk_{1} - \bk_{2} = 0$
(not a nonzero umklapp momentum $\bG$).
Thus, in the vertex amplitude~\eqref{Lambda_vertex_1} we can set ${\rm e}^{-i\bG\cdot \ba_{2}} = 1$.

This leads to Eq.~\eqref{sigma_c} in the main text.

\section{Infrared divergence of the self-energy at $b = 0$}
\label{IR_divergence}

The self-energy term $\Sigma_{\rm c}$ presents singularities originating from the region of
small integration momenta.
Starting from Eq.~\eqref{sigma_c} we can extract the singular parts by analyzing the
expression in the region $\bk_{1} \ll 1$.
(The region $\bk_{2} \ll 1$, for finite $\bk$ gives an equal contribution to the
singularity; thus the total divergence can be obtained by multiplying by a factor 2 the
singularity originating from the $\bk_{1} \ll 1$ region).

In the limit $\bk_{1} \ll 1$, the vertex functions reduce to
\begin{equation} \label{Lambda_vertices}
\begin{split}
& \Lambda_{\lambda \lambda_{1} \lambda_{2}}^{\alpha \lambda \mu}(\bk, -\bk_{1}, -\bk_{2})
\Lambda_{\lambda_{1} \lambda_{2} \lambda'}^{\nu \rho \beta}(\bk_{1}, \bk_{2}, -\bk)  \\
& \simeq  \Lambda_{\lambda \lambda_{1} \lambda_{2}}^{\alpha \lambda \mu}(\bk, 0,
-\bk) \Lambda_{\lambda_{1} \lambda_{2} \lambda'}^{\nu \rho \beta}(0, \bk, -\bk) \\
& = -\frac{V}{4} \lt(1 - \frac{b}{3}\rt) (\lambda \lambda_{1} \lambda_{2} - 1) (\lambda'
\lambda_{1} \lambda_{2} - 1) \\
& \times \lt(|f_{\bk}| (\lambda \delta^{\alpha 1} \delta^{\lambda \mu} + \lambda_{2}
\delta^{\mu 1} \delta^{\alpha \lambda}) + 3 \lambda_{1} \delta^{\lambda 1} \delta^{\mu
\alpha}\rt) \\
& \times \lt(|f_{\bk}| (\lambda' \delta^{\beta 1} \delta^{\nu \rho} + \lambda_{2}
\delta^{\rho 1} \delta^{\beta \nu}) + 3 \lambda_{1} \delta^{\nu 1} \delta^{\rho
\beta}\rt)
\end{split}
\end{equation}
up to terms which do not contribute to the divergent part of the integral.
The factor $(\lambda \lambda_{1} \lambda_{2} - 1) (\lambda' \lambda_{1} \lambda_{2} - 1)
$ vanishes unless $\lambda = \lambda'$; this implies that the divergent part is diagonal
in the branch index $\lambda$.
In addition, this factor fixes $\lambda_{2} = -\lambda \lambda_{1}$.

The numerator in the propagator of particle ``1'' can similarly be approximated in the
long-wavelength region as:
\begin{equation}
\begin{split}
& N_{\lambda_{1}}^{\lambda \nu}(\bk_{1}) - \epsilon_{\lambda_{1}}(\bk_{1})
\sigma_{y}^{\lambda \nu}\simeq N_{\lambda_{1}}^{\lambda \nu}(0)\\
& \simeq (3 + \lambda_{1} b) \delta^{\lambda \nu} + (3 -  b) \lambda_{1}
\sigma_{z}^{\lambda \nu}~.
\end{split}
\end{equation}

For $\lambda_{1} = -1$, when the soft boson is of the Goldstone type,
$N_{\lambda_{1}}^{\lambda \nu}(0) = 2(3 - b) \delta^{\lambda 2} \delta^{\nu 2}$.
Contracting with the vertices~\eqref{Lambda_vertices} and inserting the result in
Eq.~\eqref{sigma_c} we then find
\begin{equation}\label{sigma_c_div}
\begin{split}
& \Sigma_{\lambda}^{({\rm div}, {\rm Goldstone})}(\bk, \omega) \simeq \frac{V (3 -
b)^{3}|f_{\bk}|^{2}}{6 \epsilon_{\lambda}(\bk)}\int_{|\bk_{1}| \ll 1} \frac{{\rm d}^{2}
k_{1}}{\Omega} \\
& \frac{\sigma_{x}^{\alpha \mu} (N_{\lambda}^{\mu \rho}(\bk) -
\epsilon_{\lambda}(\bk) \sigma_{y}^{\mu \rho})\sigma_{x}^{\rho
\beta}}{\epsilon_{-1}(\bk_{1}) [\omega - \epsilon_{-1}(\bk_{1}) - \epsilon_{\lambda}(\bk -
\bk_{1}) + i 0^{+}]}~,
\end{split}
\end{equation}
up to finite terms.
A simple power counting shows that Eq.~\eqref{sigma_c_div} is IR divergent at $\omega =
\epsilon_{\lambda}(\bk)$.
However, it can be checked that $v_{0\lambda}^{+}(\bk)\sigma_{x}^{\alpha \mu}
(N_{\lambda}^{\mu \rho}(\bk) - \epsilon_{\lambda}(\bk) \sigma_{y}^{\mu
\rho})\sigma_{x}^{\rho \beta} v_{0 \lambda}(\bk) = 0$.
Thus the divergence disappears from the correction to the energy spectrum.

Consider now the contribution of pseudo-Goldstone modes for $b = 0$.
In this case, $\lambda_{1} = +1$, $\lambda_{2} = - \lambda$, and
$N_{\lambda_{1}}^{\lambda \nu}(0) = 6 \delta^{\lambda 1} \delta^{\nu 1}$.
Then we get a term which diverges on-shell:
\begin{equation}
\begin{split}
& \Sigma_{\lambda}^{({\rm div, pseudo-Goldstone})}(\bk, \omega) \simeq \frac{27 V}{2
\epsilon_{\lambda}(\bk)} \int_{|\bk_{1}| \ll 1} \frac{{\rm d}^{2}k_{1}}{\Omega}\\
& \frac{N_{-\lambda}^{\alpha \beta}(\bk) - \epsilon_{-\lambda}(\bk) \sigma_{y}^{\alpha
\beta}}{\epsilon_{1}(\bk) \lt(\omega - \epsilon_{1}(\bk_{1}) - \epsilon_{\lambda}(\bk -
\bk_{1}) + i 0^{+}\rt)}~.
\end{split}
\end{equation}

Contracting with $v_{0\lambda}(\bk)$ now does not annihilate the expression; instead for
$b \to 0$, $v^{+}_{0\lambda}(\bk)(N_{-\lambda}^{\alpha \beta}(\bk) -
\epsilon_{-\lambda}(\bk) ) \sigma_{y}^{\alpha \beta} v_{0 \lambda}(\bk) \simeq 2
\epsilon^{2}(\bk)/3$ with $\epsilon(\bk) = \sqrt{9 - |f_{\bk}|^{2}}$.

Replacing $\epsilon_{\lambda}(\bk) = \epsilon(\bk)$ we thus obtain:
\begin{equation} \label{pseudo-Goldstone_div}
\begin{split}
 & \Sigma_{\lambda}^{({\rm div, pseudo-Goldstone})}(\bk, \omega) \approx 9 V
\epsilon(\bk)\\
& \times  \int_{|\bk_{1}| \ll 1} \frac{{\rm d}^{2}k_{1}}{\Omega}
\frac{1}{\epsilon(\bk_{1}) (\omega - \epsilon(\bk_{1}) - \epsilon(\bk - \bk_{1}) + i
0^{+})}~.
\end{split}
\end{equation}

The expression presents a singularity at any $\bk$ for $\omega = \epsilon(\bk)$.
In particular, the imaginary part of $\Sigma$ has a finite jump, and the real part a
logarithmic divergence.
Adding a small magnetic field $b$, the divergence is regularized.
For $b \ll 1$, the leading behavior can be found replacing in
Eq.~\eqref{pseudo-Goldstone_div}, $\epsilon(\bk_{1}) \to \epsilon_{1}(\bk_{1})$.
For nonzero $\bk$, we can neglect instead the dependence on $b$ of
$\epsilon(\bk - \bk_{1})$.
This can be justified by the fact that the dispersion is strongly dependent on $b$ near
$\bk = 0$, where the gap is $\simeq \sqrt{12 b}$. For nonvanishing momentum, instead, the
dispersion receives only small corrections (perturbative with respect to $b$).

Expanding for small $\bk_{1}$ we then find:
\begin{equation}
\begin{split}
 & \Sigma_{\lambda}^{({\rm div, pseudo-Goldstone})}(\bk, \omega) \approx 9 V
\epsilon(\bk)  \int_{|\bk_{1}| \ll 1} \frac{{\rm d}^{2}k_{1}}{\Omega} \\
& \frac{1}{\epsilon_{1}(\bk_{1}) (\omega - \epsilon_{1}(\bk_{1}) - \epsilon(\bk) +
(\mathbf{v}_{\bk} \cdot \bk_{1}) + i  0^{+})}~.
\end{split}
\end{equation}

The expression is convergent and purely real on shell (for $\omega = \epsilon(\bk)$).
The on-shell self-energy depends on the field as $\approx \ln 1/b$.

\section{Renormalized perturbation theory at zero physical field}
\label{renormalized_perturbation_theory}

As discussed in the main text, we analyze the spectrum at zero physical field using a renormalized
perturbation theory, including a magnetic-field and an energy-bandwidth renormalization.
To derive this approximation, we can rewrite the Hamiltonian at zero magnetic field, $H
= \sum_{\langle i, j \rangle} m^{\alpha}_{i} m^{\alpha}_{j} + \frac{V}{4} M^{\alpha \beta}M^{\gamma
\delta} m^{\alpha}_{i} m^{\beta}_{i} m^{\gamma}_{j} m^{\delta}_{j}$, by adding and subtracting a
term
\begin{equation}
\delta H = (z - 1) H + \frac{z b}{2} \sum_{i} M^{\alpha \beta} m^{\alpha}_{i} m^{\beta}_{i}~.
\end{equation}

We then consider $H + \delta H$ as the Hamiltonian and $- \delta H$ as a perturbation.
As a result, in the calculation of the self-energy diagrams, we can use $H  + \delta H$ as the
Hamiltonian.
This leads to a diagrammatic expansion in which the propagators in the internal lines are given by:
\begin{equation} \label{G0_ren}
\begin{split}
\tilde{G}_{0 \lambda}^{-1}(\bk, \omega, b, z)  = & -z \lt(3 + \frac{b}{3} \lambda |f_{\bk}|\rt)
\delta^{\alpha \beta} - \omega \sigma^{\alpha \beta}_{y} \\
 & + z \lt(1 - \frac{b}{3}\rt) \lambda |f_{\bk}| \sigma_{z}^{\alpha \beta}~.
\end{split}
\end{equation}

Note that not only the propagator, but also the vertices have to be calculated at finite $b$ in the
computation of the self-energy diagrams.
Taking into account the transformation under energy rescaling, we find that the self-energy at
one-loop order in renormalized perturbation theory, is given by $z \Sigma(\bk, \omega/z, b)$, where
$\Sigma$ is the one-loop self-energy calculated in bare perturbation theory at finite magnetic
field.

The Dyson equation at lowest order can be written as $G^{-1}(\bk, \omega) = \tilde{G}_{0
\lambda}^{-1}(\bk, \omega, b, z)- z \Sigma(\bk, \omega/z, b) - C(\bk, \omega)$, where $\tilde{G}_{0
\lambda}^{-1}(\bk, \omega, b)$ is the renormalized propagator in Eq.~\eqref{G0_ren} and $C(\bk,
\omega)$ is the correction due to the subtraction $- \delta H$.

The correction $\delta H$ is of the same order as the self-energy $\Sigma$.
Thus we can treat it in the tree-level (zero-loop approximation).
As a result, $\tilde{G}_{0 \lambda}^{-1}(\bk, \omega, b) - C(\bk, \omega)$ can be identified with
the propagator calculated at tree-level propagator in the original model (with Hamiltonian $H = H +
\delta H - \delta H$).

We arrive therefore at the Dyson equation
\begin{equation}\label{Dyson_ren}
G^{-1}(\bk, \omega) = G_{0}^{-1}(\bk, \omega) - z\Sigma(\bk, \omega/z, b)~,
\end{equation}
where $G_{0 \lambda \lambda'}^{-1}(\bk, \omega) = -( 3 + \omega \sigma_{y}  - \lambda |f_{\bk}|
\sigma_{z}) \delta_{\lambda \lambda'}$ is the inverse propagator at zero field.

To calculate the spectrum we consider a perturbative solution for the zero eigenvalues and
eigenvectors of Eq.~\eqref{Dyson_ren}.
A direct application of perturbation theory to Eq.~\eqref{Dyson_ren} would proceed by studying
$\Sigma$ as a perturbation to the zero eigenvectors of $G_{0}^{-1}$.
This however would lead to a problematic expression.

In fact, in this framework, the spectrum at zero order is $\omega = \sqrt{9 - |f_{\bk}|^{2}}$ and
the eigenvectors are
\begin{equation}
v_{0\lambda}(\bk, 0) = \frac{1}{\sqrt{6}} \frac{i \sqrt{3 + \lambda |f_{\bk}|}}{\sqrt{3 - \lambda
|f_{\bk}|}}~.
\end{equation}

At first order, it can then be shown that the energy shift is given by the condition that the $2
\times 2$ matrix
\begin{equation}
B_{\lambda \lambda'} = - v_{0 \lambda}^{+}(\bk, 0) \lt( \Delta \omega \sigma_{y}  - z
\Sigma(\bk, \omega/z) \rt)v_{0 \lambda'}(\bk)
\end{equation}
has a zero eigenvalue.
This equation however involves the matrix elements $v_{0 \lambda}^{+}(\bk, 0) \Sigma(\bk, \omega/z)
v_{0 \lambda}(\bk, 0)$, with the vectors $v_{0}(\bk, 0)$ calculated \emph{at zero field}.
The analysis in App.~\ref{IR_divergence} on the other hand implies that the divergences due to
Goldstone modes cancel when we consider the matrix element $v_{0 \lambda}^{+}(\bk, b) \Sigma(\bk,
\omega/z, b) v_{0 \lambda}(\bk, b)$ along the vector $v_{0 \lambda}(\bk, b)$ computed at finite
field $b$ (the same field appearing in the self-energy).
Thus the expression $v_{0 \lambda}^{+}(\bk, 0) \Sigma(\bk, \omega/z, b) v_{0 \lambda}(\bk, 0)$ is
IR-divergent.

This problem can be resolved in the following way, consistent with the renormalization discussed
above.
Instead of expanding in powers of $\Sigma$, we can expand in the difference $G^{-1}-
\tilde{G}^{-1} = G_{0}^{-1} - \tilde{G}^{-1} - \Sigma $ with $\tilde{G}^{-1}$ the renormalized
propagator in Eq.~\eqref{Dyson_ren}.
If the renormalized perturbation theory is approximately self-consistent, then $G^{-1} \approx
\tilde{G}_{0}^{-1}$, and this difference should be smaller than $G^{-1} - G_{0}^{-1} = -\Sigma$.

Using this procedure, we can solve the equations to determine the spectrum, $\det G^{-1}(\bk,
\omega)$, using $\tilde{G}^{-1}(\bk, \omega) = 0$ as the zero-order starting point.
The leading-order solution is then given by the zero-order eigenvalues and eigenvectors of
$\tilde{G}^{-1}$, which are given by $\omega = z \epsilon_{\lambda}(\bk, b)$ and
\begin{equation}
v_{0 \lambda}(\bk, b) = \frac{1}{\sqrt{2 \lt(3 + \frac{b}{3} \lambda |f_{\bk}|\rt)}} \begin{pmatrix}
i \sqrt{3 + \lambda |f_{\bk}|} \\
\sqrt{3 - \lt(1 - \frac{2}{3} b\rt) \lambda |f_{\bk}|}
                               \end{pmatrix}
\end{equation}

It is simple to check that $\det \tilde{G}^{-1}(\bk, z \epsilon_{\lambda}(\bk, b)) = 0$ and
$\tilde{G}^{-1}(\bk, z \epsilon_{\lambda}(\bk, b)) v_{0 \lambda}(\bk, b) = 0$.
These are the same vectors which were introduced in Eqs.~\eqref{w_vectors} and~\eqref{v} of the main
text.
At first order in $\omega - z \epsilon_{\lambda}(\bk, b)$, the energy shift can be by imposing the
vanishing of the diagonal matrix element
\begin{equation}
v_{0\lambda}^{+}(\bk, b) \lt[ G^{-1}(\bk, \omega) - \tilde{G}^{-1}(\bk, z \epsilon_{\lambda}(\bk,
\omega)) \rt] v_{0 \lambda}(\bk, b) = 0~.
\end{equation}

Using $\tilde{G}^{-1}(\bk, z \epsilon_{\lambda}(\bk, b)) v_{0 \lambda}(\bk, b) = 0$ and replacing
the on-shell approximation $\omega = z \epsilon_{\lambda}(\bk)$ in the argument of the
self-energy we find the expression for the spectrum
\begin{equation} \label{spectrum_ren_0}
v_{0\lambda}^{+}(\bk, b) \lt[ G_{0}^{-1}(\bk, \omega, 0) - z \Sigma(\bk, \epsilon_{\lambda}(\bk, b)
\rt] v_{0
\lambda}(\bk, b) = 0~.
\end{equation}

This can be calculated using the explicit expressions for $v_{0\lambda}(\bk, b)$.
We find:
\begin{equation} \label{spectrum_ren_1_order}
\begin{split}
& -\lt(3 + z \Sigma_{\lambda}(\bk) \rt) \lt(3 + \frac{b}{3} \lambda |f_{\bk}| \rt) +
\epsilon_{\lambda}(\bk, b) \omega \\
& \qquad + \lt(1 - \frac{b}{3}\rt) |f_{\bk}|^{2} = 0~,
\end{split}
\end{equation}
where $\Sigma_{\lambda}(\bk) = v_{0 \lambda}^{+}(\bk, b) \Sigma(\bk, \epsilon_{\lambda}(\bk, b))
v_{0 \lambda}(\bk, b)$.

The expression now depends only on IR-convergent matrix elements.
Eq.~\eqref{spectrum_ren_1_order} is equivalent to~\eqref{spectrum} in the main text.

In order to discuss the role of off-diagonal matrix elements (hybridization of the two branches) we
consider for simplicity the approximation of promoting Eq.~\eqref{spectrum_ren_0} to a $2 \times 2$
matrix equation.
In particular we consider the equation $\det B (\bk, \omega) = 0$ with
\begin{equation} \label{matrix_eq}
\begin{split}
&B_{\lambda \lambda'}(\bk, \omega) =  v_{0\lambda}^{+}(\bk, b) \Big[ G_{0}^{-1}(\bk, \omega) \\
& \qquad - z \Sigma (\bk, e_{\lambda \lambda'}(\bk), b) \Big] v_{0 \lambda'}(\bk,
b) = 0~,
\end{split}
\end{equation}
with $e_{\lambda \lambda'}(\bk) = \delta_{\lambda \lambda'} \epsilon_{\lambda}(\bk) + (1 -
\delta_{\lambda \lambda'}) \epsilon(\bk, 0)$.

By an explicit calculation we find:
\begin{equation}
\begin{split}
& B_{\lambda \lambda'}(\bk, \omega) =  \Bigg[-3 + \frac{\epsilon_{\lambda}(\bk, b) \omega}{3 +
\frac{b}{3}\lambda |f_{\bk}|} \\
& \qquad + \frac{\lt(1 - \frac{b}{3}\rt) |f_{\bk}|^{2}}{3 + \frac{b}{3}\lambda |f_{\bk}|}\Bigg]
\delta_{\lambda \lambda'} - z \Sigma_{\lambda \lambda'}(\bk, e_{\lambda \lambda'}(\bk), b)~.
\end{split}
\end{equation}

The equation $\det B = 0$ can be recast equivalently as
\begin{equation}
\begin{split}
& \det \Bigg[\omega \delta_{\lambda \lambda'} - \frac{9 - |f_{\bk}|^{2} + \frac{b}{3} (3 + \lambda
|f_{\bk}|)\lambda |f_{\bk}|}{\epsilon_{\lambda}(\bk)} \delta_{\lambda \lambda'} \\
& - z\sqrt{\frac{\lt(3 + \frac{b}{3} \lambda |f_{\bk}|\rt) \lt(3
+ \frac{b}{3} \lambda' |f_{\bk}|\rt)}{\epsilon_{\lambda}(\bk, b) \epsilon_{\lambda'}(\bk, b)}}
\Sigma_{\lambda \lambda'}(\bk, e_{\lambda \lambda'}(\bk))\Bigg] = 0~.
\end{split}
\end{equation}

Thus, the spectrum is determined by the eigenvalues of the matrix $C_{\lambda \lambda'}(\bk)$
defined in the main text (Eq.~\eqref{C_k}).

\subsection*{Solution of the one-loop self-consistency relations}

The self-consistency relations~\eqref{ren_1} and~\eqref{ren_2} are simple to solve because in the
first-order approximation considered here $\Sigma$ is proportional to $V$.
By a simple algebra we find:
\begin{equation}\label{z_A}
z = \frac{1 - A(b)}{1 - 2 A(b)}~,
\end{equation}
where
\begin{equation} \label{A_b}
A(b) = \frac{2b{\rm Re}\Sigma_{\lambda}(Y, b)}{(3 + b) {\rm Re} \Sigma_{+1}(0, b)}~.
\end{equation}

The ratio does not depend explicitly on $V$, because the factor $V$ in the self-energies cancels in
the numerator and the denominator of Eq.~\eqref{A_b}.
Thus, Eq.~\eqref{z_A} is sufficient to deduce the bandwidth renormalization as a function of
the renormalized field $b$ (using a numerical integration of ${\rm Re}\Sigma_{\lambda}(Y, b)/V$ and
${\rm Re} \Sigma_{+1}(0, b)/V$).

The value of the coupling $V$ can then be deduced replacing the solution into Eq.~\eqref{ren_1}.
This gives:
\begin{equation} \label{V}
V = \frac{(2 z -1)6 b}{z (3 + b) \Sigma_{+1}(0, b)/V}~.
\end{equation}

Since $\Sigma\propto V$, the ratio $\Sigma_{+1}(0, b)/V$ does not depend explicitly on $V$.
Thus, Eq.~\eqref{V} gives directly $V$ as a function of the renormalized field $b$.

\section{Longitudinal spin correlation function}
\label{a_2mag}

After the replacement $S^{z}_{i} = S - n_{i} = S - M^{\alpha \beta} m^{\alpha}_{i}
m^{\beta}_{i}/2$ and the separation $m^{\alpha}_{i} = \langle m^{\alpha}_{i} \rangle + \delta
m_{i}^{\alpha}$, the spectral function of longitudinal spin fluctuations,
\begin{equation}
\begin{split}
{\cal S}^{zz}(\bq, \omega) & = \frac{1}{N} \sum_{i, j} \int_{-\infty}^{\infty} {\rm
d}t~{\rm e}^{i \omega t - i \bq \cdot (\bx_{i} - \bx_{j})} \langle S^{z}_{i}(t) S^{z}_{j}(0)
\rangle ~,
\end{split}
\end{equation}
can be splitted into different terms.

Here, we consider for simplicity only the contribution coming from the correlations $\langle \delta
m_{i} \delta m_{i} \delta m_{j} \delta m_{j} \rangle$.
These correspond to density fluctuations involving only non-condensed bosons.

In the calculation, we make the approximation of considering only the lowest-order loop diagram.
At zero temperature, and at finite $\bq$ and $\omega$ we obtain:
\begin{equation}
\begin{split}
{\cal S}_{\rm n.c.}^{zz}(\bq, \omega)  & = \frac{-1}{N} \sum_{i, j} {\rm Re}
\Bigg\{\int_{0}^{\infty} {\rm
d}t~{\rm e}^{i \omega t - i \bq \cdot (\bx_{i} - \bx_{j})}
\\ & G_{ij}^{\alpha \beta}(t) G_{ij}^{\alpha \beta}(t)\Bigg\}~.
\end{split}
\end{equation}

We find:
\begin{equation} \label{S_nc}
\begin{split}
& {\cal S}_{\rm n.c.}^{zz}(\bq, \omega)  = \frac{\pi}{24} \sum_{\lambda_{1},
\lambda_{2}} \int_{\rm BZ}   \frac{{\rm d}^{2}k_{1}}{\Omega}  \Bigg\{\frac{ \delta(\omega -
\epsilon_{\lambda_{1}}(\bk_{1}) - \epsilon_{\lambda_{2}}(\bk_{2}))}{\epsilon_{\lambda_{1}}(\bk_{1})
\epsilon_{\lambda_{2}}(\bk_{2})} \\
& \times {\rm Tr} \lt[(N_{\lambda_{1}}(\bk_{1}) -
\epsilon_{\lambda_{1}}(\bk_{1}))(N_{\lambda_{2}}(\bk_{2}) +
\epsilon_{\lambda_{2}}(\bk_{2}))\rt] \\
& \times \lt(1 + \frac{\lambda_{1} \lambda_{2} {\rm Re} \lt[f_{\bk_{1}}
f_{\bk_{2}}\rt]}{|f_{\bk_{1}} f_{\bk_{2}}|}\rt)\Bigg\}~.
\end{split}
\end{equation}

To compute ${\cal S}^{zz}_{\rm n.c.}(\bq, \omega)$ numerically we replace the energy-conserving
delta function with a broadened Gaussian.
To generate Fig.~\ref{2mag} we used an energy-broadening of $0.05$ and performed integrations by
discretizing the Brillouin zone on a grid of $600 \times 600$ $\bk$ points.

\end{document}